\documentclass[aps,pra,10pt,twocolumn,showpacs,superscriptaddress,preprintnumbers]{revtex4-1}

\usepackage[utf8]{inputenc}  % UTF-8 text encoding: üöä
\usepackage[T1]{fontenc}
\usepackage[english]{babel}  % english language

\usepackage{graphicx}% Include figure files
\usepackage{bm}% bold math

\usepackage{amsmath,amssymb,amscd,amsthm} % commutative diagrams, theorems

\usepackage{color}
\usepackage{verbatim} % block comments
\usepackage{xspace}
\usepackage{boxedminipage}
\usepackage{rotating}

\usepackage[bookmarks=false,pdfstartview={FitH}]{hyperref}

\theoremstyle{plain}
\newtheorem{theorem}{Theorem}%[section]
\newtheorem{corollary}{Corollary}%[section]
\theoremstyle{definition}
\newtheorem{example}{Example}%[section]
\renewcommand{\Re}{\operatorname{Re}}
\newcommand{\unity}{\ensuremath{{\rm 1 \negthickspace l}{}}}

\newcommand{\trace}{\operatorname{tr}}
\newcommand{\diag}{\operatorname{diag}}

\DeclareMathOperator{\VEC}{vec}

\newcommand{\ket}[1]{\ensuremath{| #1 \rangle}{}}

\newcommand{\ketbra}[2]{\ensuremath{| #1 \rangle \langle #2 |}{}}

\newcommand{\expt}[1]{\ensuremath{\langle #1 \rangle}{}}

\newcommand{\Partial}[2]{\ensuremath \frac{\partial{#1}}{\partial{#2}}{}}

\newcommand{\su}{\mathfrak{su}}

\newcommand{\MSC}{\ensuremath{\sf{MSC}}\xspace}
\newcommand{\MMC}{\ensuremath{\sf{MMC}}\xspace}
\newcommand{\KSC}{\ensuremath{\sf{KSC}}\xspace}
\newcommand{\KMC}{\ensuremath{\sf{KMC}}\xspace}
\newcommand{\DSC}{\ensuremath{\sf{DSC}}\xspace}

\newcommand{\reach}{\mathfrak{reach}}
\newcommand{\pos}{\mathfrak{pos}}
\newcommand{\grape}{{\sc grape}\xspace}
\newcommand{\dynamo}{{\sc dynamo}\xspace}

\newcommand{\be}{\begin{equation}}
\newcommand{\ee}{\end{equation}}

\newcommand{\expfactorbitflip}{\epsilon}
\newcommand{\expfactoramp}{\epsilon}

\begin{document}
%\bibliographystyle{prsty}

%\preprint{special issue {\em Quantum Control Theory for Coherence and Information Dynamics}}

\title{How to Transfer between Arbitrary $n$-Qubit Quantum States by Coherent Control and Simplest Switchable Noise on a Single Qubit}
%\title{Arbitrary $n$-qubit state transfer using coherent control and switchable single-qubit noise}

\author{Ville Bergholm}
\email{ville.bergholm@iki.fi}
\affiliation{Dept.~Chemistry, Technical University Munich (TUM), D-85747 Garching, Germany}
\affiliation{Institute for Scientific Interchange Foundation (ISI), I-10126 Turin, Italy}
\author{Thomas Schulte-Herbr{\"u}ggen}\email{tosh@ch.tum.de}
\affiliation{Dept.~Chemistry, Technical University Munich (TUM), D-85747 Garching, Germany}

\date{\today}% It is always \today, today,


\begin{abstract}
We explore reachable sets of open $n$-qubit quantum systems, the coherent parts 
of which are under full unitary control and that have just one qubit whose Markovian
noise amplitude can be modulated in time such as to provide an additional 
degree of incoherent control. In particular, adding bang-bang control of amplitude 
damping noise (non-unital) allows the dynamic system to act transitively on the entire 
set of density operators. This means one can transform any initial quantum state into 
any desired target state. Adding switchable bit-flip noise (unital), on the other hand, 
suffices to explore all states majorised by the initial state. We have extended our 
open-loop optimal control algorithm (\dynamo package) by such degrees of incoherent control so that these 
unprecedented reachable sets can systematically be exploited in experiments. 
%As exemplified for 
%an experimental setting in ion traps, switching noise can simplify the setup for state transfer from 
%measurement-based closed-loop feedback to open-loop schemes with the same capabilities. 
As illustrated for an ion trap experimental setting, open-loop control
with noise switching can accomplish all state transfers one can get by the
more complicated measurement-based closed-loop feedback schemes.
 
\end{abstract}

\maketitle
%%%%%%%%%%%%%%%%%%%%%%%%%%%%%%%%%%%%%%%%%%%%%%%%%%%%%%%%%%%%%%%%%%%
%%%%%%%%%%%%%%%%%%%%%%%%%%%%%%%%%%%%%%%%%%%%%%%%%%%%%%%%%%%%%%%%%%%
%%%%%%%%%%%%%%%%%%%%%%%%%%%%%%%%%%%%%%%%%%%%%%%%%%%%%%%%%%%%%%%%%%%

%%%

%%%%%%%%%%%%%%%%%%
%{\em Introduction.}
%%%%%%%%%%%%%%%%%%

Recently, dissipation has been exploited for quantum state
engineering~\cite{VWC09,KMP11} so that evolution under constant noise leads
to long-lived entangled fixed-point states. Lloyd and Viola~\cite{VioLloyd01} showed 
that closed-loop feedback from one {\em resettable ancilla qubit} suffices to simulate any quantum dynamics of open systems. 
Both concepts were used to combine coherent  dynamics with optical pumping
on an ancilla qubit for dissipative preparation of entangled states~\cite{BZB11} or quantum maps~\cite{SMB12}. ---
Besides, full control over the Kraus operators~\cite{Rabitz07b} 
or the environment~\cite{Pechen11} allows for interconverting arbitrary
quantum states.

Manipulating quantum systems with high precision is paramount to exploring their
properties for pioneering experiments, and also in view of new technologies~\cite{DowMil03}.
Therefore it is most desirable to extend the current toolbox of optimal control~\cite{WisMil09} by 
systematically incorporating dissipative control parameters.

Here we prove that it suffices to include as a new control parameter 
a single bang-bang switchable Markovian noise amplitude for one qubit (no ancilla) 
into an otherwise noiseless and coherently controllable network to increase the power
of the dynamic system such that any target state can be reached from any initial state.
Thus we extend our optimal-control platform {\sc dynamo}~\cite{PRA11} by controls
over Markovian noise sources.
To illustrate possible applications, we demonstrate
the initialisation step~\cite{VincCriteria} of quantum computing,
i.e.\ the transfer from the thermal state to the pure state
$\ket{00\ldots0}$ (as well as the opposite process), the
interconversion of random pairs of mixed states,
and finally the noise-driven generation of maximally entangled states.

\medskip
%%%%%%%%%%%%%%%%%%
{\em Theory.}
%%%%%%%%%%%%%%%%%%
We consider the quantum Markovian master equation
of an $n$-qubit system
as a bilinear control system~($\Sigma$):
\begin{equation}\label{eqn:master}
\dot\rho(t) = -(i\hat H_u +\Gamma)\rho(t)\quad\text{and}\quad\rho(0)=\rho_0
\end{equation}
with $H_u:= H_0 + \sum_j u_j(t) H_j$ comprising the free-evolution Hamiltonian $H_0$, the control
Hamiltonians $H_j$ switched by piecewise constant control amplitudes $u_j(t)\in\mathbb R$ and $\hat H_u$
as the corresponding commutator superoperator.
Take $\Gamma$ to be of Lindblad form
\begin{equation}\label{eqn:Lindblad}
\Gamma(\rho) := -\sum_\ell \gamma_\ell(t)\big(V_\ell\rho V_\ell^\dagger - 
	\tfrac{1}{2}(V^\dagger_\ell V_\ell \rho + \rho V^\dagger_\ell V_\ell)\big)\;,
\end{equation}
where now $\gamma_\ell(t)\in[0,\gamma_*]$ with $\gamma_*>0$ will be used as
additional piecewise constant control parameters.
% in addition to the $\{u_j(t)\}$.

In the sequel we will consider mostly systems with a single Lindblad generator.
In the {\em non-unital case} it is the Lindblad generator for {\em amplitude damping},~$V_a$,
while in the {\em unital case} it is the one for {\em bit flip},~$V_b$,  defined as
\begin{equation}\label{eqn:amp-damp+bit-flip}
V_a := \unity_2^{\otimes{n-1}}\otimes\ketbra{0}{1}\quad\text{and}\quad
V_b := \unity_2^{\otimes{n-1}}\otimes\sigma_x/2\;.
\end{equation}

Here we follow the Lie-algebraic setting along the lines of~\cite{JS72,SchiFuSol01,ZS10,Alt03,DHKS08}.
As in~\cite{DHKS08}, we say the control system on $n$ qubits meets the condition for (weak)
Hamiltonian controllability if the Lie closure under commutation is
\begin{equation}\label{eqn:closure-wh}
\hspace{-3mm}
\expt{iH_0, iH_j\,|\,j=1,\dots , m}_{\sf Lie} = \su(N)\;\text{with}\; N:=2^n .
\end{equation}
Now the {\em reachable set} $\reach_\Sigma(\rho_0)$ is defined as the set of all
states $\rho(T)$ with $T\geq 0$ that can be reached from $\rho_0$ following the dynamics of ($\Sigma$).
If Eqn.~\eqref{eqn:closure-wh} holds,
without relaxation one can steer from any initial state $\rho_0$ to
any other state $\rho_{\rm target}$ with the same eigenvalues.
In other words, for $\gamma=0$ the control system ($\Sigma$) acts transitively on the 
unitary orbit $\mathcal U(\rho_0):=\{U\rho_0 U^\dagger\,|\,U\in SU(N)\}$ 
of the respective initial state $\rho_0$. This holds for
any $\rho_0$ in the set of all density operators, termed $\pos_1$ henceforth.

Under coherent control and {\em constant noise} ($\gamma>0$ non-switchable) it is difficult to give precise
reachable sets for general $n$-qubit systems that satisfy Eqn.~\eqref{eqn:closure-wh}
only upon including the system Hamiltonian ($H_0$). Based on seminal work by Uhlmann~\cite{Uhlm71, Uhlm72, Uhlm73}, 
majorisation criteria that are powerful if $H_0$ is not needed to meet  Eqn.~\eqref{eqn:closure-wh} 
\cite{Yuan10, Yuan11} now just give upper bounds to reachable sets by inclusions. 
But with increasing number of qubits $n$, these inclusions become increasingly inaccurate
and have to be replaced by Lie-semigroup methods as described in~\cite{ODS11}.

In the presence of {\em bang-bang switchable relaxation on a single
  qubit} in an $n$-qubit system, here we show that the situation
improves significantly and one obtains 
the following results, both proven in the Supplement~\cite{SuppMat}:

\begin{theorem}\label{thm:transitivity}
Let $\Sigma_a$ be an $n$-qubit bilinear control system as in Eqn.~\eqref{eqn:master}
satisfying Eqn.~\eqref{eqn:closure-wh} for $\gamma=0$.
Suppose the $n^{\rm th}$ qubit (say) undergoes (non-unital)
amplitude-damping relaxation, the noise amplitude of which can be
switched in time between two values as $\gamma(t)\in\{0,\gamma_*\}$ with $\gamma_*>0$. 
If the free evolution Hamiltonian~$H_0$ is diagonal (e.g., Ising-$ZZ$ type),
and if there are no further sources of decoherence, then 
the system $\Sigma_a$
acts transitively on the set of all density operators $\pos_1$:
\begin{equation}
\overline{\reach_{\Sigma_a}^{\phantom{1}}(\rho_0)}=\pos_1 \quad\text{for all }\, \rho_0\in\pos_1\;,
\end{equation}
where the closure is understood as the limit $\gamma_* T\to\infty$.
\end{theorem}

\begin{theorem}\label{thm:majorisation}
Let $\Sigma_b$ be an $n$-qubit bilinear control system as in Eqn.~\eqref{eqn:master}
satisfying Eqn.~\eqref{eqn:closure-wh} ($\gamma=0$)
now with the $n^{\rm th}$ qubit (say) undergoing (unital) bit-flip relaxation
with switchable noise amplitude 
$\gamma(t)\in\{0,\gamma_*\}$.
If the free evolution Hamiltonian~$H_0$ is diagonal (e.g., Ising-$ZZ$ type),
and if there are no further sources of decoherence, then 
in the limit $\gamma_* T\to\infty$ the reachable set to $\Sigma_b$
explores all density operators majorised by the initial state $\rho_0$, i.e.
\begin{equation}
\hspace{-3mm}
\overline{\reach_{\Sigma_b}^{\phantom{1}}(\rho_0)}=\{\rho\in\pos_1 \,|\,\rho\prec\rho_0\}\;\text{for any }\, \rho_0\in\pos_1\,.
\end{equation}
\end{theorem}
The conditions for the drift Hamiltonian~$H_0$
in the theorems above can be relaxed. The details are given
in the Supplement~\cite[\ref{sec:proofs}]{SuppMat},
along with the proofs.

The scenarios of Eqn.~\eqref{eqn:amp-damp+bit-flip}
can be generalised to the Lindblad generator
$V_\theta:=\left(\begin{smallmatrix} 0 & (1-\theta)\\ \theta & 0 \end{smallmatrix}\right)$
with $\theta\in[0,1]$. 
If $\theta \neq 1/2$ the noise qubit has a unique fixed point~$\rho_\infty(\theta)$.
Comparing this to the canonical density operator of a qubit with energy level splitting~$\Delta$,
relaxation by~$V_\theta$ gives the same fixed point as
equilibrating the 
system via the noisy qubit with a local heat bath
of inverse temperature
\begin{equation*}
%\label{eqn:AlgCooling-delta:main}
\beta := \frac{1}{k_B T}
=
\frac{2}{\Delta} \operatorname{artanh}\left(\delta(\theta)\right)
\quad \text{with} \quad
\delta(\theta) := \frac{\bar\theta^2-\theta^2}{\bar\theta^2+\theta^2}
\end{equation*}
using the shorthand $\bar\theta:=1-\theta$.
As limiting cases,  pure amplitude damping is
brought about by a bath with zero temperature (i.e.~$\theta=0$), while
pure bit-flip corresponds to 
the high-temperature limit $T \to \infty$ (i.e. $\theta \to
\tfrac{1}{2}$).

While a single-qubit system with unitary control and bang-bang switchable noise generator $V_\theta$
can clearly be steered (asymptotically) to any state with purity less or equal to the
larger of the purities of $\rho_0$ and $\rho_\infty(\theta)$,
the situation for $n\geq 2$ qubits is more involved:
using coherent control,
relaxation of a diagonal initial state can be limited to a single pair of
eigenvalues at a time if all the remaining ones can be arranged in pairs
$(\rho_{ii}, \rho_{jj})$, each satisfying
% reverse the effect of the noise
\begin{equation}\label{eqn:theta-stop:main}
\theta^2 / \bar{\theta}^2 \le \rho_{ii}/\rho_{jj} \le \bar\theta^2 / \theta^2\;.
\end{equation}
Yet Eqn.~\eqref{eqn:theta-stop:main} poses no restriction in the important task of cooling:
starting from the maximally mixed state, optimal control protocols with period-wise relaxation by $V_\theta$ 
interspersed with unitary permutation of diagonal density operator elements is more general than
the partner-pairing approach~\cite{SMW05} to \emph{algorithmic cooling} with bias 
$\delta(\theta)$ and $0\leq\theta<{1}/{2}$. This type of algorithmic cooling proceeds also just on
the diagonal elements of the density operator, but it involves no transfers limited to a single pair
of eigenvalues (details in the Supplement~\cite[\ref{app:B}]{SuppMat}). 

\medskip
%%%%%%%%%%%%%%%%%%
{\em Exploring Model Systems.}
%%%%%%%%%%%%%%%%%%
To challenge our extended optimal control algorithm,
we first consider two examples of state transfer where the target states are
{\em on the boundary} of the respective reachable sets of the initial states, in other words, they can only be
reached asymptotically ($\gamma_* T\to\infty$). 
To illustrate Theorems~\ref{thm:transitivity} and~\ref{thm:majorisation},
we then demonstrate noise-driven transfer (i) between \emph{random pairs of states} under controlled
amplitude damping noise and (ii) between random pairs of states satisfying $\rho_{\rm target}\prec\rho_0$
under controlled bit-flip noise in Examples 3 and~4.
We finish with entanglement generation in an experimental trapped ion
system in Example~5.

In Examples 1--4, our system is an $n$-qubit chain with uniform
\mbox{Ising-$ZZ$} nearest-neighbour couplings given by
$H_0 := \pi J \sum_k \tfrac{1}{2} \sigma_z^{(k)} \sigma_z^{(k+1)}$
and piecewise constant $x$ and~$y$ controls (that need not be bounded) on each qubit
locally, so the control systems satisfy Eqn.~\eqref{eqn:closure-wh}.
We add controllable noise (either amplitude-damping or
bit flip) with amplitude $\gamma(t)\in[0,\gamma_*]$
acting on one terminal qubit.
In all the examples we set~$\gamma_* = 5 J$.

%%%%%%%%%%%%%%%%%%%%%%%%%%%%%%%%%%%%
\begin{figure}[Ht!]
\hspace{10mm}{\sf (a)}\hspace{34mm}\sf{(b)} $\hfill$\\[-0mm]
\includegraphics[width=0.43\columnwidth]{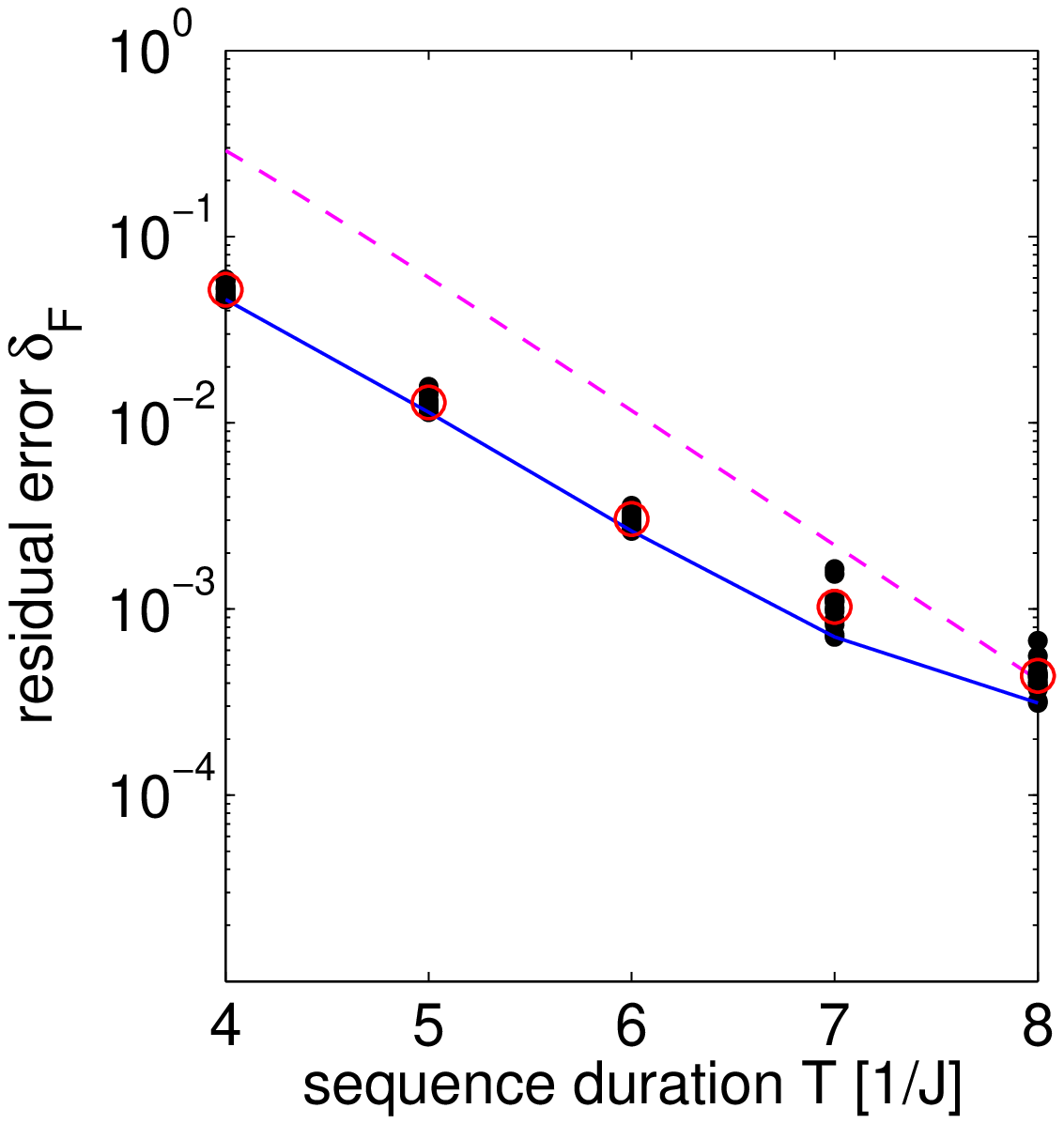}
\includegraphics[width=0.43\columnwidth]{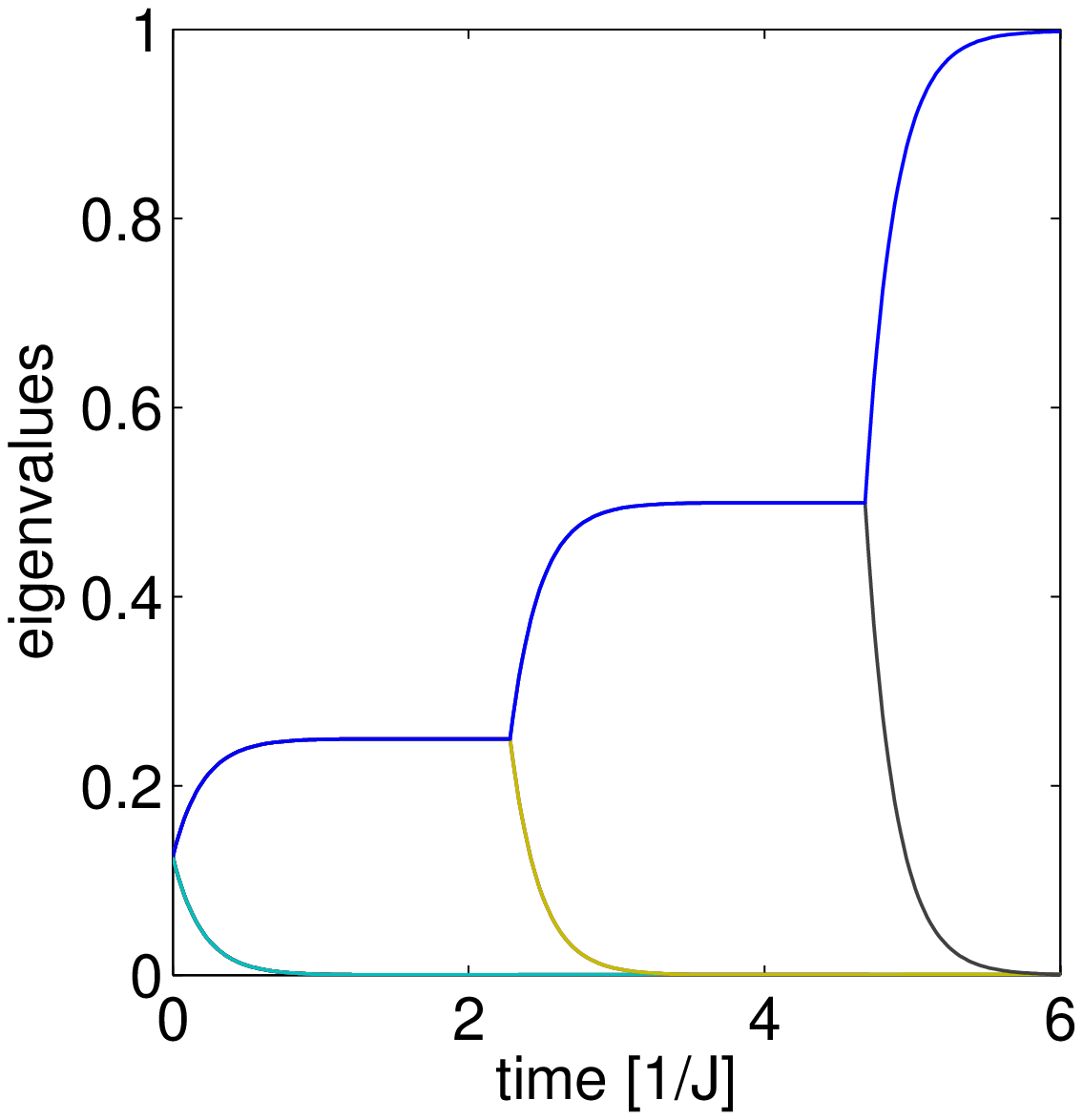}\\
\hspace{10mm}{\sf (c)}$\hfill$\\
\includegraphics[width=0.9\columnwidth]{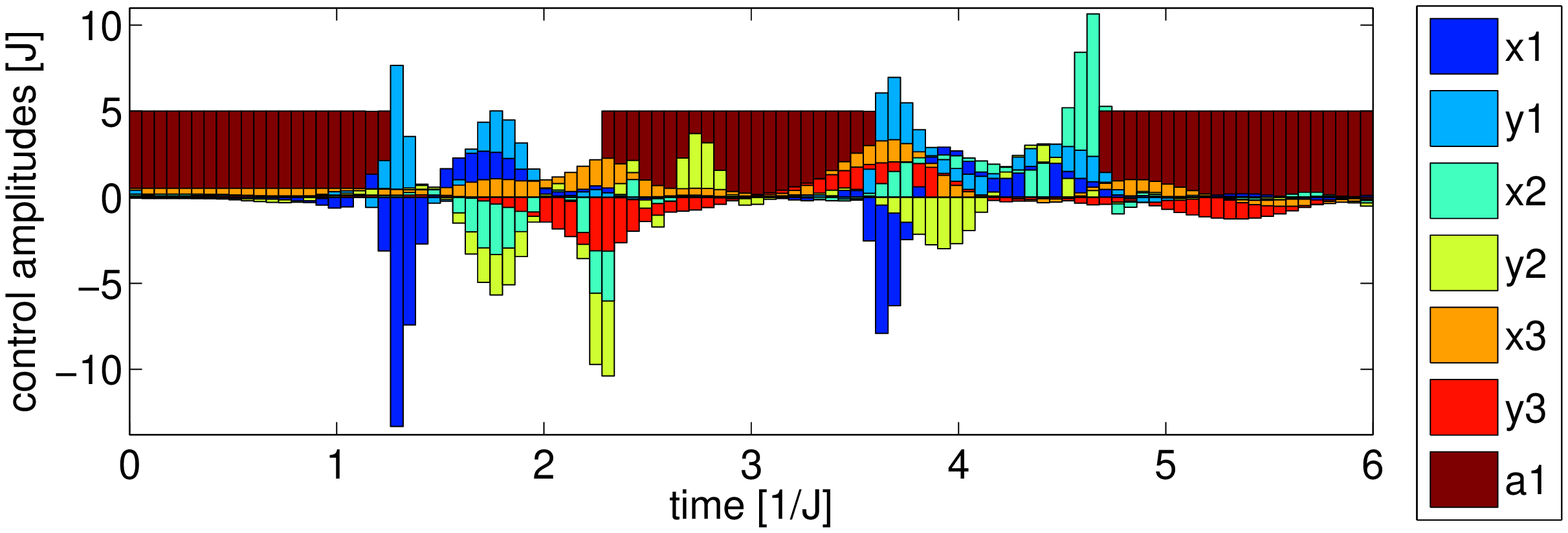}
\caption{\label{fig:th0}
Transfer
from the thermal state~$\rho_\text{th}=\frac{1}{8}\unity$ to the zero-state~$\rho_{\ket{000}}$ 
in a \mbox{3-qubit} Ising-$ZZ$ chain
with controlled amplitude-damping noise on qubit one
as in {\bf Example~1}.
(a)~Quality versus total sequence duration~$T$.
The dashed line gives the upper bound from Eqn.~\eqref{eq:time-ex1},
and the dots (red circles for averages) individual optimization runs with 
random initial sequences.
Noise amplitudes were initialised in three
distinct blocks of equal duration to help the optimisation towards an
economic solution.
(b)~Evolution of the eigenvalues under the best of the $T=6/J$
solutions.
This sequence (c) shows three relaxative periods with maximal noise
amplitude~$\gamma_{a1}$ for transforming eigenvalues, while 
unitary actions governed by ($u_{x\nu}, u_{y\nu}$) 
mainly take place in the intervals between.
Each purely unitary segment is of the approximate duration~$1/J$,
corresponding to the duration of a single $i$-swap.
%Note that unitary controls on qubit one
%$(u_{x1},u_{y1})$ are minute when the noise on qubit one is active.
}
\end{figure}
%%%%%%%%%%%%%%%%%%%%%%%%%%%%%%%%%%%%

\begin{example}%[Cooling a thermal state by amplitude damping]

Here, as for initialising a quantum computer~\cite{VincCriteria},
the task is to turn the thermal initial state
$\rho_{\text{th}}:=\tfrac{1}{2^n}\unity$ into the pure target state
$\ket{00\ldots0}$
by unitary control and controlled amplitude damping.
For $n$ qubits, the task can be accomplished in an \mbox{$n$-step} protocol: 
let the noise act on each qubit~$q$ for the time~$\tau_q$ to 
populate the state $\ketbra{0}{0}$,
and permute the qubits between the steps.
A linear chain requires $\sum_{q=1}^n (q-1)=\binom{n}{2}$ %$ =\frac{n(n-1)}{2}$ such
nearest-neighbour swaps.
Since all the intermediate states are diagonal, the swaps can be
replaced with $i$-swaps, each taking a time of~$\frac{1}{J}$
under the Ising-$ZZ$ coupling.
The residual Frobenius error~$\delta_F$
is minimised when all the $\tau_q$ are equal, giving
\be
\label{eq:th0est}
\delta_{F_a}^2(\expfactoramp)
= 1 -2 \left(1-\tfrac{\expfactoramp}{2}\right)^n +\left(1 -\expfactoramp +\tfrac{1}{2} \expfactoramp^2\right)^n,
\ee
where $\expfactoramp:=e^{-\gamma_* T_n/n}$ and~$T_n:=\sum_q\tau_q$.
Linearizing this expression and adding the time for the $i$-swaps,
the total duration~$T_a$ of this simple protocol as a function of $\delta_F$ 
amounts (in first order) to
\begin{equation}\label{eq:time-ex1}
T_a \approx \binom{n}{2}\tfrac{1}{J} +\tfrac{n}{\gamma_*} 
	\ln \Big(  \tfrac{\sqrt{n(n+1)}}{2 \delta_{F_a}^{\phantom{|}} } \Big).
\end{equation}

Fig.~\ref{fig:th0} demonstrates that optimal control can outperform
this simple scheme by parallelising part of the unitary
transfer with the amplitude-damping driven \/`cooling\/' steps.
Interestingly, the initialisation task can still be accomplished to a good
approximation when unavoidable constant dephasing noise on all the
three qubits is added, as shown in the Supplement~\cite[\ref{sec:furtherresults}]{SuppMat}.

%INTERNAL: So is there a scheme with duration
%\begin{equation}
%T' = \tfrac{n-1}{J} + \tfrac{n}{\gamma_*} 
%	\ln \Big(  \tfrac{\sqrt{n(n+1)}}{2 \delta_F^{\phantom{|}} } \Big)
%\end{equation}
%???}
\end{example}

\begin{example}%[Heating a pure state by bit flip noise]

In turn, consider \/`erasing\/' the pure initial
state~$\ket{00\ldots0}$
to the thermal state~$\rho_{\text{th}}$.
% of Eqn.~\eqref{eqn:bit-flip}
Under controlled amplitude damping,
this can be accomplished exactly, each round splitting the populations in half with a total
time of
$
T'_a = \binom{n}{2}\tfrac{1}{J} + \tfrac{n}{\gamma_*} \ln(2).
$
However, with {\em bit-flip noise} this transfer 
can only be obtained asymptotically.
One may use a similar $n$-step protocol as in the previous example,
this time approximately erasing each qubit to a state proportional
to~$\unity$.
Again, optimal control greatly outperforms this simple scheme.
Results and details are shown in the Supplement~\cite[\ref{sec:furtherresults}]{SuppMat}.

%%%%%%%%%%%%%%%%%%%%%%%%%%%%%%%%%%
\begin{figure}[Ht!]
\hspace{2mm}{\sf (a)}\hspace{25mm}{\sf (b)}\hspace{25mm}{\sf(c)}$\hfill$\\
\raisebox{-.6mm}{\includegraphics[width=0.327\columnwidth]{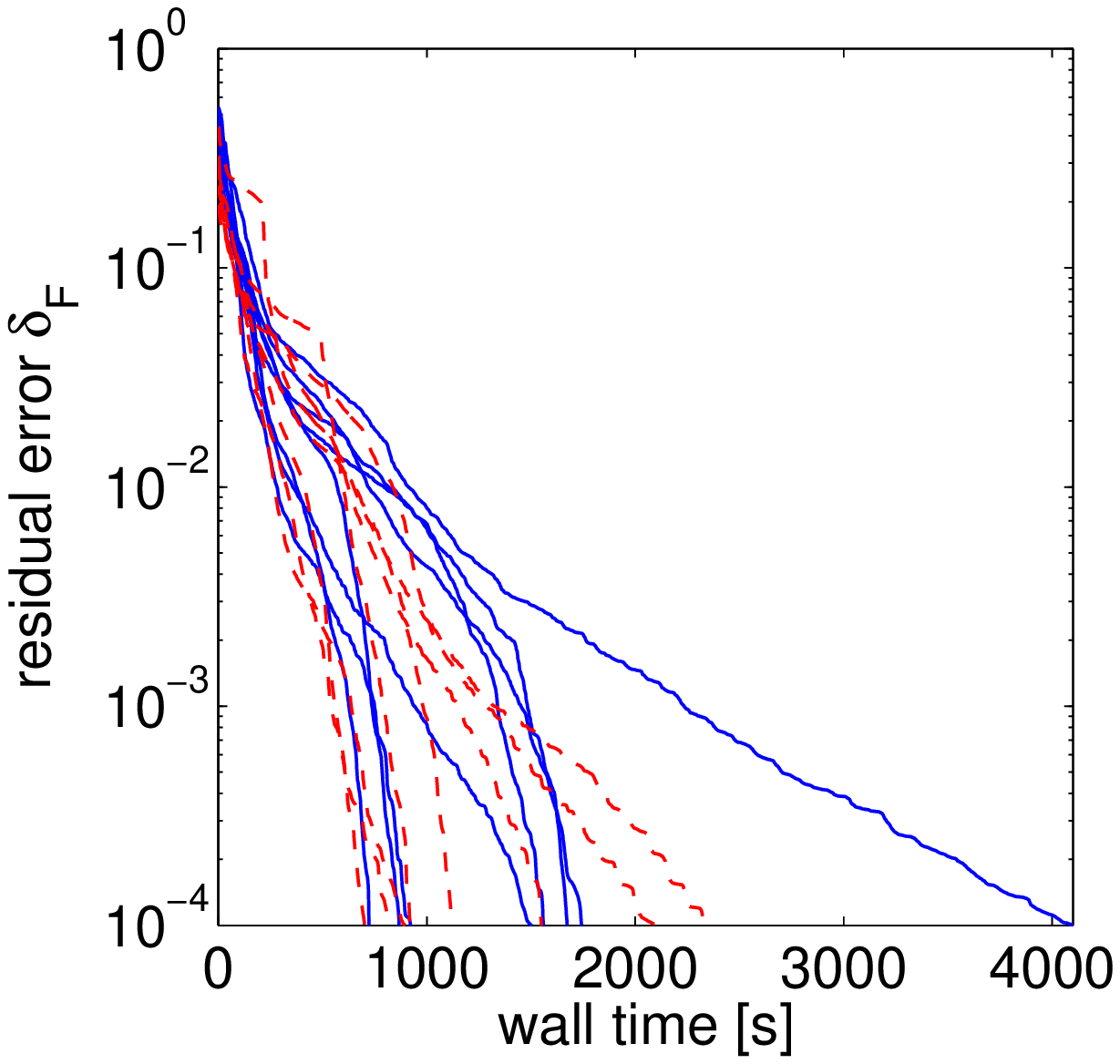}}
\includegraphics[width=0.32\columnwidth]{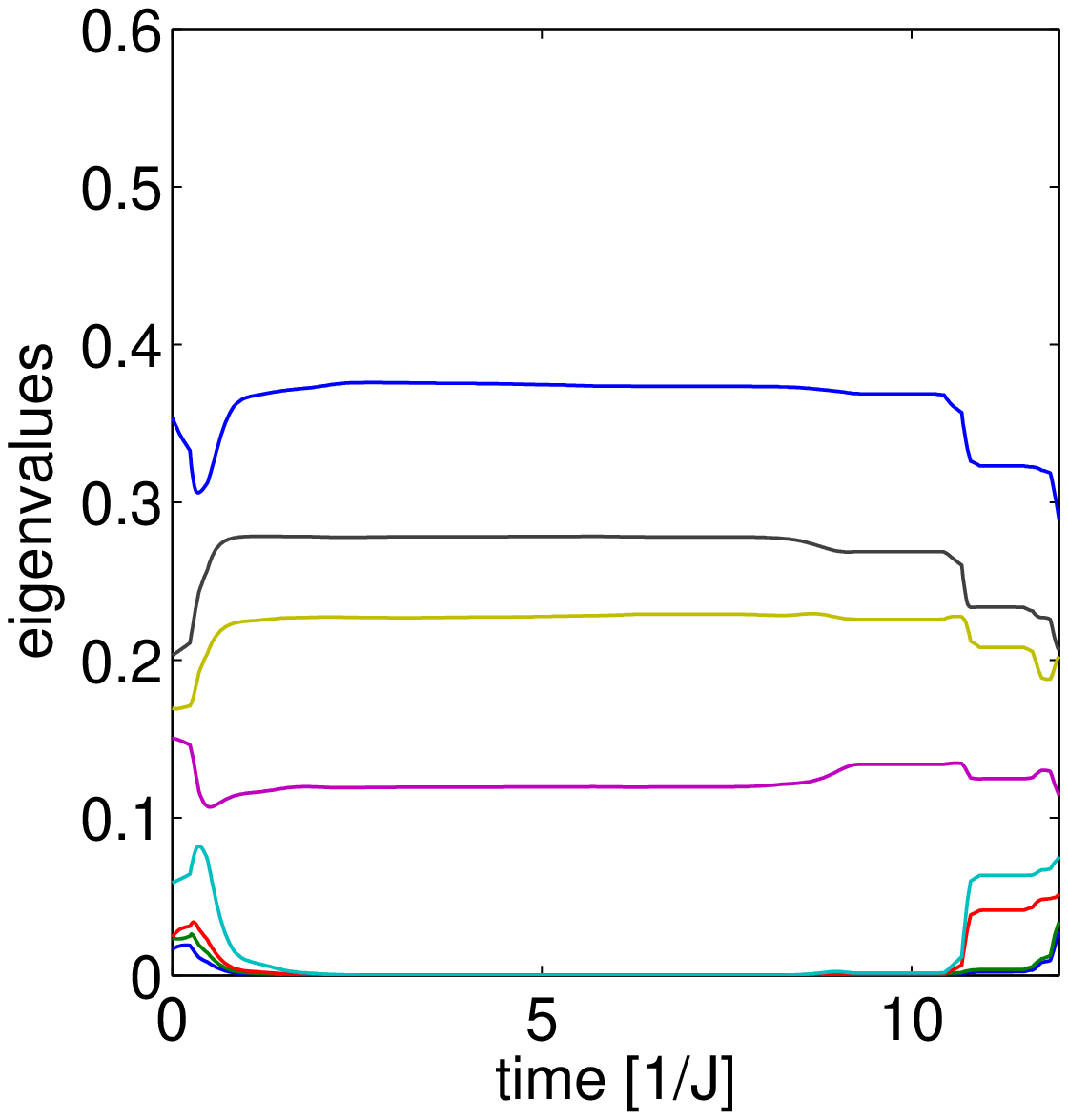}
\includegraphics[width=0.32\columnwidth]{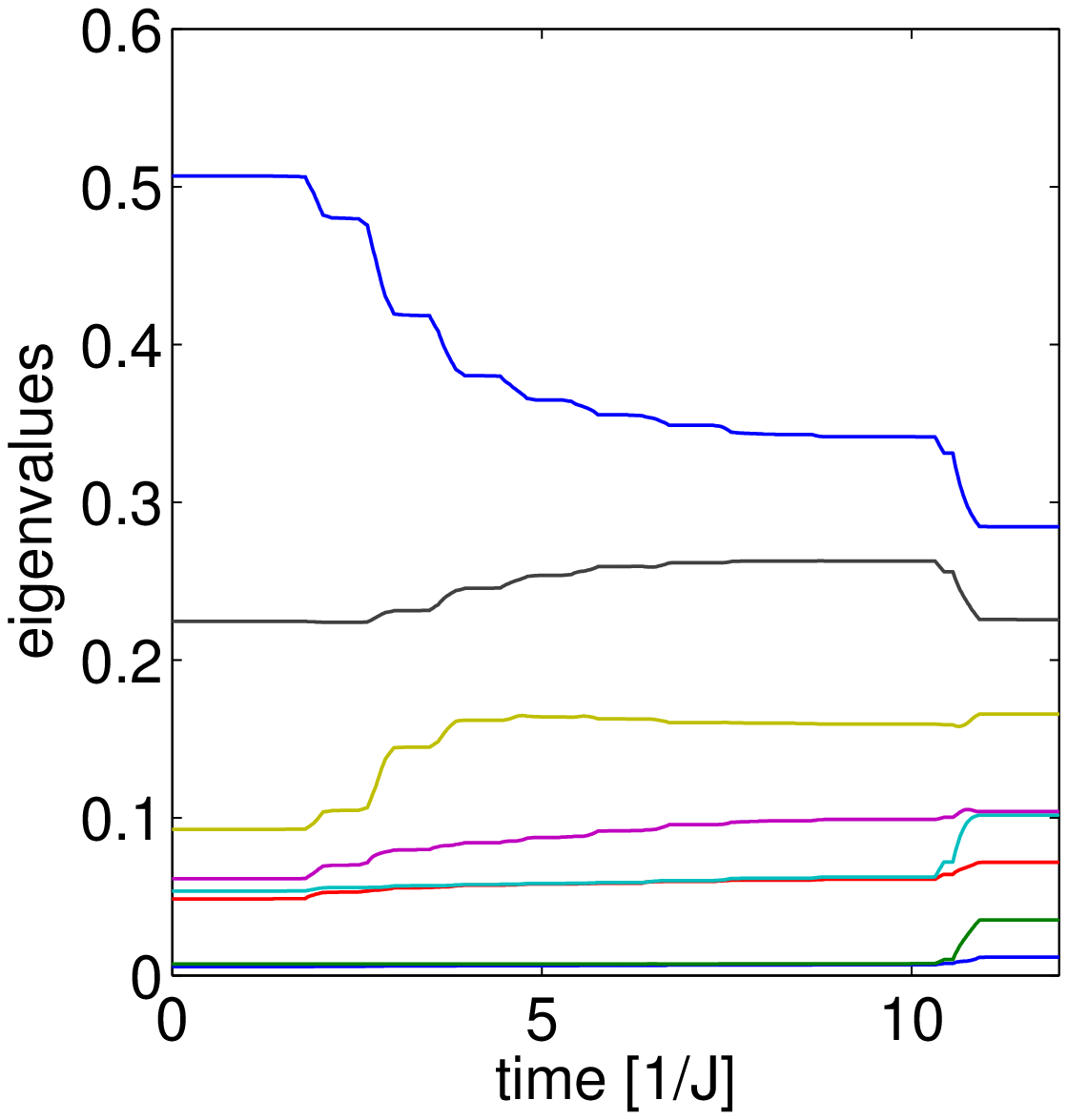}
\caption{\label{fig:rand}
(a) Quality vs.~computation time for state transfer between 
pairs ($\rho_0,\rho_{\rm target}$)
of random \mbox{3-qubit} states using controlled amplitude-damping
noise (solid) in addition to local unitary control. 
Same for random pairs  ($\rho_0,\rho_{\rm target}$)
with $\rho_{\rm target}\prec\rho_0$
under controlled bit flip noise (dashed).
In both cases (a) shows the median of 9
optimisation runs for each of the eight random state pairs.
Representative examples of evolution of the eigenvalues for an amplitude-damping transfer (b) and
for a bit-flip transfer (c). In the former case, a typical feature is the
initial zeroing of the smaller half of the eigenvalues while the larger
half are re-distributed among themselves. Only at the very end are the
smaller eigenvalues resurrected.
}
\end{figure}
%%%%%%%%%%%%%%%%%%%%%%%%%%%%%%%%%%

\end{example}

\begin{example}%[Random state transfer with switchable amplitude damping plus unitary control]
We illustrate transitivity under
controlled amplitude damping on one qubit plus general unitary control
by transfers between pairs of random \mbox{3-qubit} density operators.
Fig.~\ref{fig:rand}(a) shows the algorithm to converge well to~$\delta_F = 10^{-4}$. 
%Only for one pair (in a total of eight) we see no converging median
%within the 1800 seconds allowed out of 9 runs in total
%(but even in that case we have some converging runs).
As shown in Fig.~\ref{fig:rand}(b), the best sequences
seem to zero the smaller half of the eigenvalues as soon as possible
just to revive them in the very end after the larger half has been balanced among themselves.
\end{example}

%\bigskip
\begin{example}%[Random state transfer with bit flip noise]
Similarly allowing for controlled bit-flip noise on one qubit plus
general unitary control,
we address the transfer between arbitrary pairs of 3-qubit density
operators %($\rho_0,\rho_{\rm target}$) 
with  $\rho_{\rm target}\prec\rho_0$.
Fig.~\ref{fig:rand}(a) again illustrates the good convergence of the algorithm.
Many of these profiles exhibit ``terraces'' which could indicate local
quasi-optima.
The unital case may be harder to optimise in general:
(1)~the majorisation condition entails that
a suboptimal transfer made early in the sequence cannot be outbalanced later 
in the control sequence; it can only be mended in a next iteration; 
(2)~the necessity for simultaneous decoupling 
(like Trotterisation in the proof of Thm.~\ref{thm:majorisation})
adds to the hardness of the optimisation.
\end{example}

%%%%%%%%%%%%%%%%%%%%%%%%%%%%%%%%%%%
\begin{figure}[Ht!]
\hspace{1.5mm}{\sf (a)}$\hfill$\\
\includegraphics[width=0.93\columnwidth]{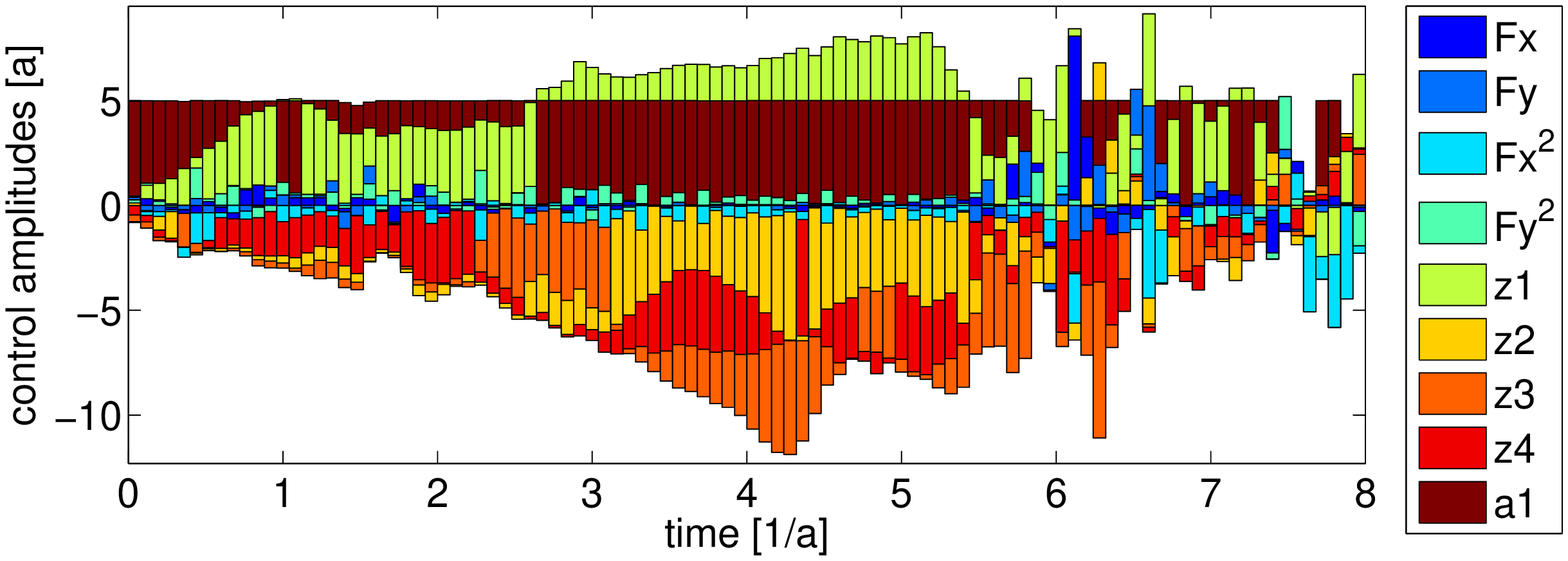}\\
\hspace{1.5mm}{\sf (b)\hspace{55mm} (c)}$\hfill$\\
\hspace{-0.5mm}\includegraphics[width=0.75\columnwidth]{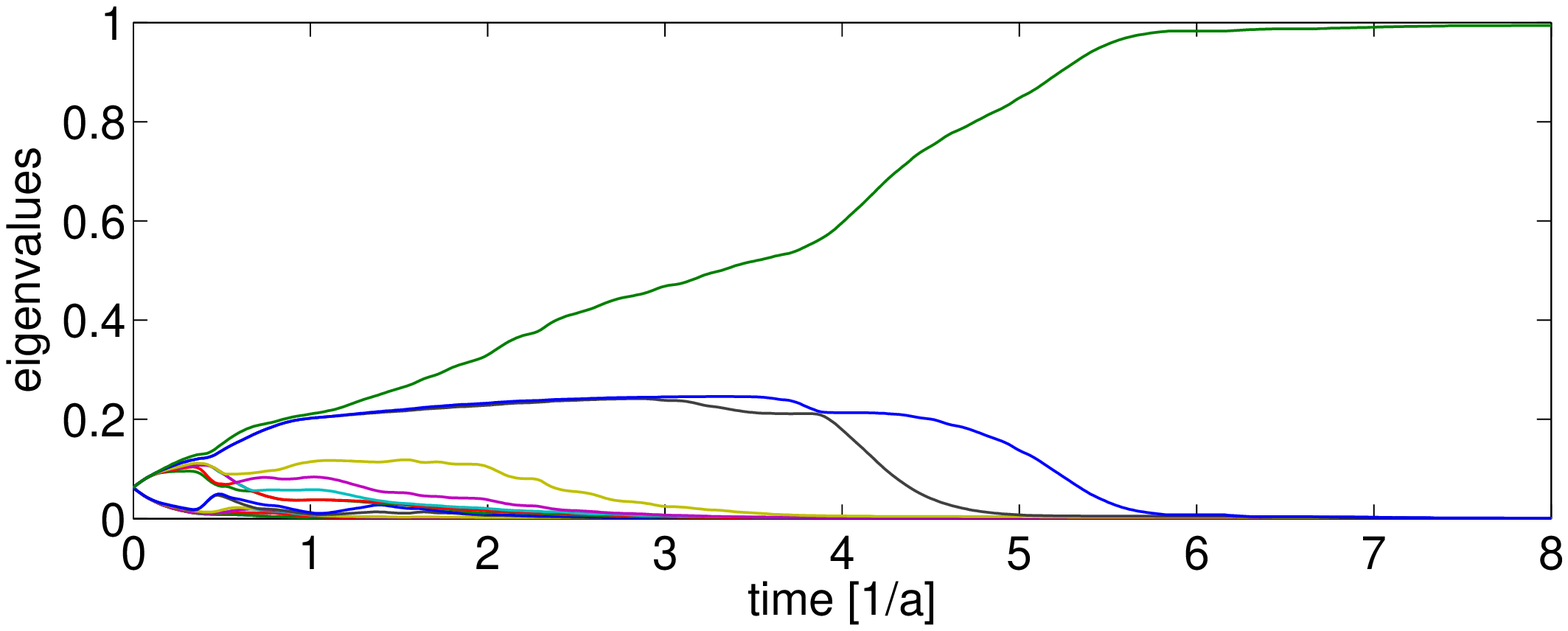}
\raisebox{7mm}{\hspace{-5mm}\includegraphics[width=0.28\columnwidth]{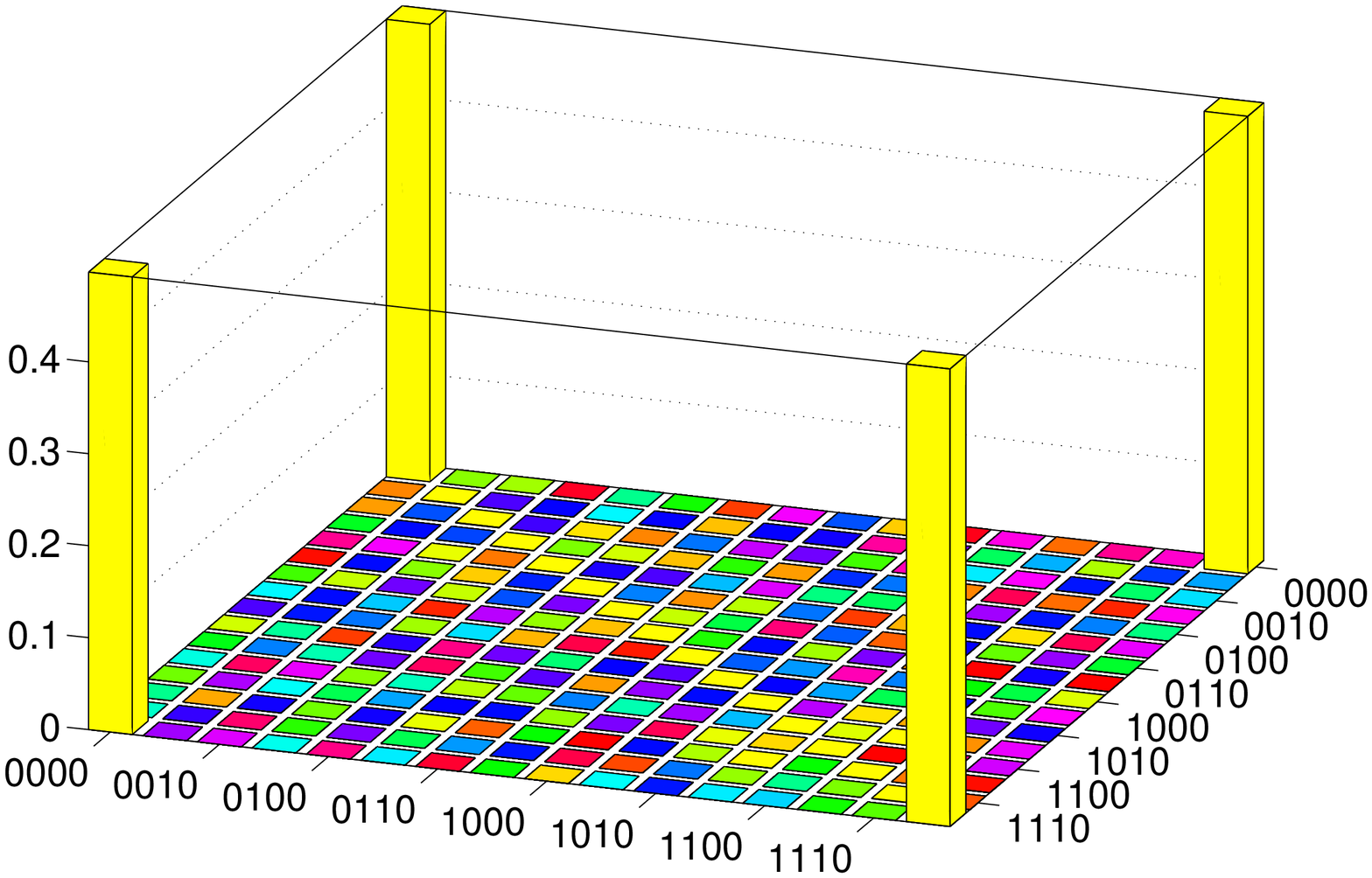}}\\[-8mm]
\raisebox{7mm}{\hspace{59mm}\includegraphics[width=0.28\columnwidth]{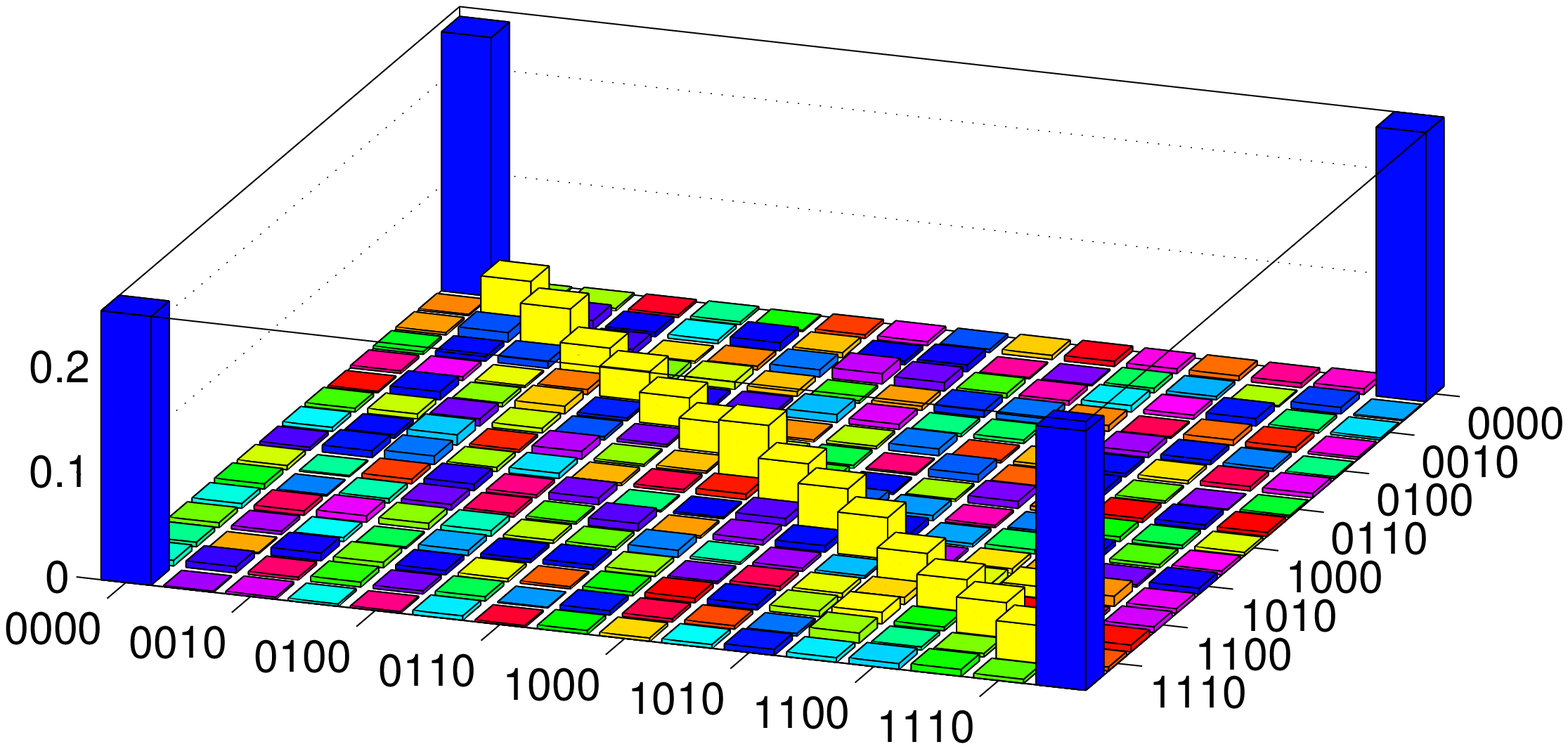}}\\[-8mm]
\raisebox{59mm}{\scriptsize \hspace{52mm} error $\times 100$}\\[-59mm]
\caption{\label{fig:blatt}
State transfer from the thermal state to the four-qubit GHZ state in the
ion-trap system of {\bf Example~5} similar to~\cite{BZB11}. By
controlled \/`pumping\/' (amplitude damping) on one qubit, one
can do without closed-loop measurement-based circuits involving an additional ancilla qubit
as required in~\cite{BZB11,VioLloyd01}.
Our  sequence (a) drives the system
to the state (c), which differs from the target state $\ket{\text{GHZ}_4}$
by an error of $\delta_F\simeq 5\cdot10^{-3}$.
The time evolution of the eigenvalues (b) illustrates parallel action
on all the eigenvalues under the sequence.
}
\end{figure}
%%%%%%%%%%%%%%%%%%%%%%%%%%%%%%%%%%%

%\bigskip
\begin{example}
The final example addresses entanglement generation in a system
similar to the one in~\cite{BZB11}. It consists of
four trapped ion qubits coherently controlled by lasers. 
On top of individual local $z$-controls ($u_{z1},\dots,u_{z4}$) on each qubit, one can
pulse on all the qubits simultaneously by the joint $x$ and $y$-controls
$F_\nu:=\tfrac{1}{2}\sum_{j=1}^4 \sigma_{\nu j}$ with $\nu=x,y$ as well as by
the quadratic terms $F^2_\nu:=(F_\nu)^2$. All the control amplitudes are expressed
as multiples of an interaction strength $a$.
In contrast to~\cite{BZB11}, where the protocol resorts to an ancilla qubit to be 
added  (following~\cite{VioLloyd01}) for a {\em measurement-based
circuit on the $4+1$ system}, here we do {\em without the ancilla qubit} by
making just the terminal qubit subject to controlled amplitude-damping 
noise with strength $\gamma_{a1}$,
to drive the system from the thermal initial state
$\rho_{\text{th}}:=\tfrac{1}{2^n}\unity$ to the pure entangled target state
$\ket{\text{GHZ}_4} = \tfrac{1}{\sqrt{2}}(\ket{0000}+\ket{1111})$.
As shown in Fig.~\ref{fig:blatt}, the
optimised controls use the noise with maximal amplitude over its entire
duration interrupted just by two short periods of purely unitary control.
%The target state is
%reached with a residual error as small as $\delta_F \approx 5\cdot10^{-3}$.
%Fig.~\ref{fig:blatt}(c) shows the resulting state and the
%difference from the target.
\end{example}

%\medskip

%%%%%%%%%%%%%%%%%%
{\em Discussion.}
%%%%%%%%%%%%%%%%%%
By unitary controllability, we may diagonalise the initial and the target states.
So transferring a diagonal initial state into a diagonal target state 
can be considered as the {\em normal form of the state-transfer problem}. It can be treated analytically, because
it is easy to separate dissipation-driven {\em changes of eigenvalues} from unitary coherent
actions of {\em permuting eigenvalues} and decoupling drift Hamiltonians. 
Now the difference between optimising amplitude-damping non-unital transfer (as in Thm.~\ref{thm:transitivity})
and bit-flip unital transfer (as in Thm.~\ref{thm:majorisation}) becomes evident:
In the {\em non-unital case}, transitive action on the set of all density operators clearly helps to escape from 
suboptimal intermediate control sequences during the optimisation.
Yet in {\em the unital case}, the majorisation condition 
$\rho_{\rm target}\prec\rho(t)\prec\rho_0$ for all $0\leq t\leq T$
and the boundary conditions $\rho(0)=\rho_0$, $\rho(T)=\rho_{\rm target}$ (at worst for \mbox{$\gamma_* T\to\infty$})
explain potential algorithmic traps: 
one may easily arrive at an intermediate state $\rho_m(t)\prec\rho_0$ that comes closer
to the target state, but will never reach it as it fails to meet
the reachability condition $\rho_{\rm target}\prec\rho_m(t)$;
see the Supplement~\cite[\ref{sec:HLP}]{SuppMat} for how to avoid this.

%\begin{comment}
%For analysing whether local optima have entailed traps in the case of  switchable {\em unital} noise (which
%we have not observed in the cases studied here), 
%it is easy to devise a formal Lagrange-type penalty parameter $\mu:=\sum_k \mu(t_k)$  summing the 
%violation of the majorisation condition in each time interval
%\begin{equation}
%\mu(t_k) := -\sum\limits_{\nu=1}^{N-1} \Big(\sum\limits_{i=1}^{\nu} y_i(t_k) - \sum\limits_{i=1}^{\nu}  x_i \Big)\Big|_{\text{if $<0$}}\;,
%\end{equation}
%where $ y(t_k)$ denotes the eigenvalues of $\rho(t_k)$ sorted in descending magnitude
%(similarly $x$  for $\rho_{\rm target}$). Unfortunately, since there seems to be no efficient way of differentiating it,
%it can serve for evaluating failures rather than incorporating it into the algorithm for avoiding traps.
%\end{comment}

One may contrast our method with the closed-loop control method
in~\cite{VioLloyd01} originally designed for quantum-map synthesis
using projective measurement of a coupled resettable ancilla qubit
plus full unitary control to enact
arbitrary quantum operations (including state transfers),
with Markovian evolution as the infinitesimal limit.
%Assuming that the initial state of the system is known,
Applied to state transfer, the present method instead relies on a switchable local Markovian noise source
and requires {\em no measurement nor an ancilla} \footnote{
The Supplement~\cite[\ref{sec:controllabilities}]{SuppMat}
explains how
for {\em state transfer}, Markovian open-loop controllability already implies full 
state controllability (including transfers by non-Markovian processes),
while for the lift to {\em Kraus-map controllability} it remains an open question, whether
closed-loop feedback control is not only sufficient (as established in \cite{VioLloyd01}), but
also necessary in the sense that it could not be replaced by open-loop unitary control plus
control over local Markovian noise.
}.

\medskip
%%%%%%%%%%%%%%%%%
{\em Conclusions and Outlook.}
%%%%%%%%%%%%%%%%%
We have proven that by adding as a new control parameter 
bang-bang switchable Markovian noise on just one system qubit,
an otherwise coherently controllable $n$-qubit network can explore unprecedented
reachable sets: in the case of amplitude-damping noise
(or any noise process
in its unitary equivalence class, with compatible drift) %[NOTE: $H_0$ needs to match the noise],
one can convert
{\em any} initial state $\rho_0$ into {\em any} target state $\rho_{\rm target}$,
while under switchable bit-flip noise
(or any noise process unitarily equivalent)
one can transfer any~$\rho_0$ into
any target~$\rho_{\rm target}\prec\rho_0$ {\em majorised by the initial state}.
These results have been further generalised and compared to equilibrating the system with
a finite-temperature bath. 

To our knowledge, this is the first time these features have been systematically explored
as {\em open-loop} control problems and solved 
in a minimal setting
by coherent  local controls 
and bang-bang modulation of a single local noise source that is exactly Markovian.
For {\em state transfer}, our open-loop protocol ensures full state controllability,
so here it is as powerful as the closed-loop
measurement-based feedback scheme in \cite{VioLloyd01}
(see the Supplement~\cite[\ref{sec:controllabilities}]{SuppMat}). Thus it may serve to 
simplify many experimental implementations.

We have extended our optimal-control platform {\sc dynamo}~\cite{PRA11} by controls
over Markovian noise amplitudes. Our method was also shown to 
supersede algorithmic cooling~\cite{SMW05}.
As possible applications, we demonstrated
the initialisation step of quantum computing (i.e.\ the transfer from the thermal state to the
pure zero-state~\cite{VincCriteria}), the erasure, and the interconversion of random pairs of mixed states,
as well as the noise-controlled generation of maximally entangled states.

We anticipate that our approach of switching or even modulating the amplitudes
of standard Markovian noise processes as additional open-loop control parameters (in an otherwise coherently controllable 
system) will pave the way to many experimental applications.
For instance, bit flips may be induced by
external random processes %or gradients \cite{VioHavel01},
and amplitude damping may be mimicked by pumping.
If needed to facilitate experimental implementation, 
our algorithm can be specialised such as to separate dissipative and unitary evolution.
Otherwise, the algorithm parallelises coherent and incoherent controls to an extent
usually going beyond analytical tractability.
%
%Fixed noise and tunable coupling?
%

{\em Acknowledgements.}
%%%%%%%
We wish to thank Fernando Pastawski, Lorenza Viola, and Alexander Pechen for useful comments
mainly on the relation to their works.
This research was supported by the {\sc eu} project {\sc q-essence}, the
exchange with {\sc coquit},
by {\em Deutsche Forschungsgemeinschaft} in {\sc sfb}~631 and
{\sc for}~1482,
and by the excellence network of Bavaria ({\sc enb}) through {\sc qccc}.

%%B%%%
%%%%%
%\bibliography{control21ville}

%%%%%
%%B%%%

\onecolumngrid
\appendix

\clearpage
\newpage
%%%%%%%%%%%%%%%%%%%%%%%%%%%%%%%%%%%%%%%%%%
%%%%%%%%%%%%%%%%%%%%%%%%%%%%%%%%%%%%%%%%%%
\appendix
%%%%%%%%%%%%%%%%%%%%%%%%%%%%%%%%%%%%%%%%%%
%%%%%%%%%%%%%%%%%%%%%%%%%%%%%%%%%%%%%%%%%%

\begin{center}
{\Large\bf Supplementary Material}
\end{center}

\section{Proofs of the Main Theorems and Generalisations}\label{sec:proofs}

\setcounter{theorem}{0}
\begin{theorem}
%\label{thm:transitivity}
Let $\Sigma_a$ be an $n$-qubit bilinear control system as in Eqn.~\eqref{eqn:master}
satisfying Eqn.~\eqref{eqn:closure-wh} for $\gamma=0$.
Suppose the $n^{\rm th}$ qubit (say) undergoes (non-unital)
amplitude-damping relaxation, the noise amplitude of which can be
switched in time between two values as $\gamma(t)\in\{0,\gamma_*\}$ with $\gamma_*>0$. 
If the free evolution Hamiltonian~$H_0$ is diagonal (e.g., Ising-$ZZ$ type),
and if there are no further sources of decoherence, then 
the system $\Sigma_a$
acts transitively on the set of all density operators $\pos_1$:
\begin{equation}
\overline{\reach_{\Sigma_a}^{\phantom{1}}(\rho_0)}=\pos_1 \quad\text{for all }\, \rho_0\in\pos_1\;,
\end{equation}
where the closure is understood as the limit $\gamma_* T\to\infty$.

\begin{proof}%[{\bf Proof}]
We keep the proofs largely constructive. By unitary controllability, $\rho_0$
may be chosen diagonal as $\rho_0~=:~\diag(r_0)$. 
Since a diagonal $\rho_0$ commutes with a diagonal free evolution Hamiltonian $H_{0}$, the
evolution under noise and coupling remains purely diagonal, following
\begin{equation}\label{eqn:thm1}
r(t) = \left[\unity_2^{\otimes(n-1)}\otimes\begin{pmatrix}
		1 & 1-\expfactoramp\\
		0 & \expfactoramp
\end{pmatrix}\right]  r_0\; =: R_a(t)\, r_0,
\end{equation}
where $\expfactoramp := e^{-\gamma_* t}$ and $R_a(t)$ is by
construction a stochastic matrix.
With the noise switched off, full unitary control includes arbitrary permutations
of the diagonal elements. Any of the pairwise relaxative transfers between
diagonal elements $\rho_{ii}$ and $\rho_{jj}$  (with $i\neq j$)
lasting a total time of $\tau$ can be neutralised by permuting $\rho_{ii}$ and $\rho_{jj}$ 
after a time
\begin{equation}\label{eqn:switch-time1}
\tau_{ij}:=\tfrac{1}{\gamma_*} \ln \left(\frac{\rho_{ii}e^{+\gamma_* \tau}+\rho_{jj}}{\rho_{ii}+\rho_{jj}}\right)
\end{equation}
and letting the system evolve under noise again for the remaining time $\tau-\tau_{ij}$.
Thus with $2^{n-1}-1$ such switches all but one desired transfer can
be neutralised.
As $\rho(t)$ remains diagonal under all permutations, 
relaxative and coupling processes, one can obtain any state of the form
\begin{equation}
\rho(t)=
\diag(\ldots, \: {[\rho_{ii} + \rho_{jj}\cdot (1-e^{-\gamma_* t})]}_{ii},
 \: \ldots, {[\rho_{jj}\cdot e^{-\gamma_* t}]}_{jj}, \: \ldots).
\end{equation}
Sequences of such transfers between single pairs of eigenvalues 
$\rho_{ii}$ and $\rho_{jj}$  and their permutations then generate (for
$\gamma_* T\to \infty$) the entire set of all 
diagonal density operators $\Delta\subset\pos_1$. 
By unitary controllability one gets all the unitary orbits $\mathcal U(\Delta)=\pos_1$.
Hence the result.
\end{proof}
\end{theorem}

\begin{theorem}
%\label{thm:majorisation}
Let $\Sigma_b$ be an $n$-qubit bilinear control system as in Eqn.~\eqref{eqn:master}
satisfying Eqn.~\eqref{eqn:closure-wh} ($\gamma=0$)
now with the $n^{\rm th}$ qubit (say) undergoing (unital) bit-flip relaxation
with switchable noise amplitude 
$\gamma(t)\in\{0,\gamma_*\}$.
If the free evolution Hamiltonian~$H_0$ is diagonal (e.g., Ising-$ZZ$ type),
and if there are no further sources of decoherence, then 
in the limit $\gamma_* T\to\infty$ the reachable set to $\Sigma_b$
explores all density operators majorised by the initial state $\rho_0$, i.e.
\begin{equation}
\overline{\reach_{\Sigma_b}^{\phantom{1}}(\rho_0)}=\{\rho\in\pos_1 \,|\,\rho\prec\rho_0\}\;\text{for any }\, \rho_0\in\pos_1\,.
\end{equation}

\begin{proof}
Again consider the initial state $\rho_0=:\diag(r_0)$. 
The evolution under the noise remains diagonal following
\begin{equation}\label{eqn:thm2}
r(t) = \left[\unity_2^{\otimes(n-1)}\otimes\tfrac{1}{2}\begin{pmatrix}
		(1 + \expfactorbitflip)& (1-\expfactorbitflip)  \\
		(1-\expfactorbitflip) &  (1 + \expfactorbitflip)
\end{pmatrix}\right]  r_0=: R_b(t) \, r_0,
\end{equation}
where $\expfactorbitflip:= e^{-\tfrac{\gamma_*}{2}t}$ and $R_b(t)$ is doubly stochastic.
In order to limit the relaxative averaging to the first two eigenvalues,
first conjugate $\rho_0$ with the unitary 
\begin{equation}\label{eqn:protect}
U_{12}:= \unity_2 \oplus \tfrac{1}{\sqrt{2}}
\begin{pmatrix}
  1 & -1\\
  1 & 1
\end{pmatrix}^{\oplus 2^{n-1}-1}
\end{equation}
to obtain $\rho_0':=U_{12}\rho_0 U^\dagger_{12}$.
Then the relaxation acts as a {$T$-transform}~\footnote{
	A $T$-transform is a convex combination  
	%%% WE KEEP MARSHALL-OLKIN CONVENTION THOUGH LESS NICE %%%
	$\lambda\unity+(1-\lambda) Q$, where $Q$~is a pair transposition
	matrix and $\lambda\in [0,1]$.}
on the first two eigenvalues of $\rho_0'$, while leaving the remaining ones invariant. 

Yet the protected subspaces have to be decoupled from the
free evolution Hamiltonian $H_0$ assumed diagonal. Any such $H_0$
decomposes as
$H_0 =: H_{0,1} \otimes \unity +H_{0,2} \otimes \sigma_z$,
where $H_{0,1}$ and $H_{0,2}$ are diagonal.
The term with~$H_{0,1}$
commutes with~$\rho'_0$ 
%and the bit-flip noise generator
and can thus be neglected.
The other term can be 
sign-inverted 
using $\pi$-pulses in the $x$-direction on the noisy qubit (generated by $H_{1x}$),
\begin{equation}
e^{i\pi \hat H_{1x}} e^{-t(\Gamma+i\hat H_{0,2}\otimes \sigma_z)}
e^{-i\pi \hat H_{1x}} = e^{-t(\Gamma-i\hat H_{0,2}\otimes \sigma_z)},
\end{equation}
which also leave the bit-flip noise generator invariant.
Thus $H_0$ may be fully decoupled in the Trotter limit
\begin{equation}\label{eqn:Trotter-dec}
\lim_{k\to\infty}\;(e^{-\tfrac{t}{2k}(\Gamma+i\hat H_{0,2}\otimes \sigma_z)}  e^{-\tfrac{t}{2k}(\Gamma-i\hat H_{0,2}\otimes \sigma_z)})^k =  e^{-t\Gamma}\;.
\end{equation}
By combining permutations of diagonal elements
with selective pairwise averaging by relaxation, any \mbox{$T$-transform} of $\rho_0$~\footnote{
	$R_b(t)$ of Eqn.~\eqref{eqn:thm2} covers $\lambda\in [\tfrac{1}{2},1]$, while
	$\lambda\in [0,\tfrac{1}{2}]$ is obtained by unitarily swapping the elements
	before applying $R_b(t)$;
	$\lambda=\tfrac{1}{2}$ is obtained in the limit 
	$\expfactorbitflip\to 0$. 
	}
can be obtained in the limit $\gamma_* T\to\infty$:
\begin{align}\label{eqn:T-trafo}
\notag
\rho(t) =
\diag\big(\ldots, \: &\tfrac{1}{2}[\rho_{ii} + \rho_{jj} + (\rho_{ii} - \rho_{jj})\cdot e^{-\tfrac{\gamma_*}{2}t}]_{ii}, \: \ldots,\\
&\tfrac{1}{2}[\rho_{ii} + \rho_{jj} + (\rho_{jj}-\rho_{ii}) \cdot e^{-\tfrac{\gamma_*}{2}t}]_{jj}, \: \dots \big).
\end{align}

Now recall that a vector $y\in\mathbb R^N$ majorises a vector 
$x\in\mathbb R^N$, $x\prec y$,
if and only if there is a doubly stochastic matrix $D$ with $x=D y$, where $D$ is a product of at most $N-1$ such
$T$-transforms (e.g., Thm.~B.6 in~\cite{MarshallOlkin} or Thm.~II.1.10 in~\cite{Bhatia}).
Actually, by the work of Hardy, Littlewood, and P{\'o}lya~\cite{HLP34} this sequence of 
\mbox{$T$-transforms} is constructive~\cite[p32]{MarshallOlkin} as will be made use of later.
Thus in the limit $\gamma_* T\to\infty$
all diagonal vectors $r\prec r_0$ can be reached and hence by unitary controllability 
all the states $\rho\prec\rho_0$.

Finally, to see that one cannot go beyond the states majorised by the initial state,
observe that controlled unitary dynamics combined with bit-flip relaxation is still completely positive,
trace-preserving and unital. Thus it takes the generalised form of a {\em doubly-stochastic linear map} 
$\Phi$ in the sense of~Thm.~7.1 in~\cite{Ando89}, which for any hermitian matrix~$A$ ensures 
$\Phi(A)\prec A$. Hence the (closure of the) reachable set is indeed confined to $\rho\prec\rho_0$. 
\end{proof}
\end{theorem}

\medskip
The conditions for the drift Hamiltonian~$H_0$
in theorems above can be relaxed to the following generalisations.
Any free evolution Hamiltonian~$H_0$ may be diagonalised by a unitary transformation:
$H_0 = U H_0^{\text{diag}} U^\dagger$.
The same transformation~$U$, when applied to the Lindblad
generator~$V$, yields a new Lindblad generator $V' := U V U^\dagger$.
If~$V$ satisfies Theorem~\ref{thm:transitivity} or \ref{thm:majorisation} with any diagonal free
evolution Hamiltonian, then $V'$ will satisfy them with~$H_0$.
Degenerate eigenvalues of~$H_0$ yield some freedom in
choosing~$U$ which, together with arbitrary permutations,
can be used to make~$V'$ simpler to implement (e.g. local).

\medskip
Moreover, 
the theorems above are stated under very mild conditions. So 
\begin{enumerate}
\item the theorems  hold {\em a forteriori} if the noise amplitude
	is not only a bang-bang control $\gamma(t)\in\{0,\gamma_*\}$, but may vary in time within the entire 
	interval $\gamma(t)\in[0,\gamma_*]$;  the use of this will demonstrated in more complicated systems elsewhere;
\item likewise, if several qubits come with switchable noise of the same type (unital or non-unital),
	then the (closures of the) reachable sets themselves do not alter, yet the control problems can be solved more efficiently;
\item needless to say,
	a single switchable non-unital noise process (equivalent to amplitude damping) 
	on top of unital ones suffices to make the system act transitively;
\item for systems with non-unital switchable noise (equivalent to amplitude damping),
	the (closure of the) reachable set under non-Markovian conditions 
	cannot grow, since it already encompasses the entire set of density operators 
	(see Sec.~\ref{sec:controllabilities})---yet again the control problems may become easier to solve efficiently; 
\item likewise in the unital case, the reachable set does not grow under non- Markovian conditions, since the
 	Markovian scenario already explores all interconversions
        obeying the majorisation condition (see also Sec.~\ref{sec:controllabilities});
\item the same arguments hold for a {\em coded logical subspace} that is unitarily fully controllable and coupled
	to a single physical qubit undergoing switchable noise.
\end{enumerate}

\section{Generalisation of Noise Generators and Their Relation to Coupling to Finite-Temperature Baths}\label{app:B}
\subsection{Generalised Lindblad Terms}

The noise scenarios of the previous theorems can be
generalised by using the Lindblad generator
$V_\theta:=\left(\begin{smallmatrix} 0 & (1-\theta)\\ \theta & 0 \end{smallmatrix}\right)$
with $\theta\in[0,1]$. Using the short-hand $\bar\theta:=1-\theta$, the
Lindbladian and its exponential turn into
\begin{equation}\label{eqn:GLtheta}
\Gamma(\theta) = -\begin{pmatrix} -\theta^2 & 0 & 0 &\bar\theta^2\\
						0 &\bar\theta\theta - \tfrac{1}{2} & \bar\theta\theta & 0\\
						0 &\bar\theta\theta & \bar\theta\theta - \tfrac{1}{2} & 0\\
						\theta^2 & 0 & 0 & -\bar\theta^2 \end{pmatrix},
\quad \text{and}
\end{equation}
\begin{equation}\label{eqn:expGLtheta}
e^{-\gamma_*t \,\Gamma(\theta)} =
\begin{pmatrix}
c_\theta(\bar\theta^2 + \theta^2  \varepsilon_\theta) & 0 & 0 & c_\theta\bar\theta^2(1-\varepsilon_\theta)\\
0 & \varepsilon_\theta'\cosh(\gamma_*t\bar\theta\theta) & \varepsilon_\theta'\sinh(\gamma_*t\bar\theta\theta) & 0\\
0 & \varepsilon_\theta'\sinh(\gamma_*t\bar\theta\theta) & \varepsilon_\theta' \cosh(\gamma_*t\bar\theta\theta) &0\\
c_\theta\theta^2(1-\varepsilon_\theta) & 0 & 0 & c_\theta(\theta^2 + \bar\theta^2  \varepsilon_\theta)
\end{pmatrix}
\end{equation}
with $c_\theta:=\tfrac{1}{\bar\theta^2+\theta^{2^{\phantom{|}}} }$, $\varepsilon_\theta:=e^{-\gamma_*t/c_\theta}$ 
and $\varepsilon'_\theta:= e^{\gamma_*t(\bar\theta\theta-1/2)}$ as
further short-hands.
Choosing the initial state diagonal, the action on a diagonal vector of an $n$-qubit state
takes the following form that can be decomposed into a (scaled) convex sum of
a pure amplitude-damping part and a pure bit-flip part (cp.~Eqs.~\eqref{eqn:thm1},\eqref{eqn:thm2})
\begin{equation}\hspace{-7mm}
\begin{split}
R_\theta(t) &= \unity_2^{\otimes (n-1)} \otimes \left[c_\theta\begin{pmatrix} (\bar\theta^2 + \theta^2  \varepsilon_\theta) &\bar\theta^2(1-\varepsilon_\theta)\\
						\theta^2(1-\varepsilon_\theta) &  (\theta^2 + \bar\theta^2  \varepsilon_\theta) \end{pmatrix}\right]\\[2mm]
	&= \unity_2^{\otimes (n-1)} \otimes \left[\tfrac{\bar\theta^2-\theta^2}{\bar\theta^2+\theta^2}\begin{pmatrix} 1 & (1-\varepsilon_\theta)\\
											  0 &  \varepsilon_\theta\end{pmatrix} +
		\tfrac{\theta^2}{\bar\theta^2+\theta^2}\begin{pmatrix} (1+\varepsilon_\theta) & (1-\varepsilon_\theta)\\
											(1-\varepsilon_\theta) & (1+\varepsilon_\theta)\end{pmatrix}\right]\;.
\end{split}
\end{equation}

In order to limit the entire dissipative action over some fixed time $\tau$ to the first two eigenvalues (as in Thm.~1), 
one may switch again as in Eqn.~\eqref{eqn:switch-time1} after a time
\begin{equation}
\tau_{ij}(\theta):=\frac{c_\theta}{\gamma_*}  \;
	\ln \left(\frac{e^{\gamma_*\tau/c_\theta}(\bar\theta^2\rho_{ii} - \theta^2\rho_{jj})+(\bar\theta^2\rho_{jj}-\theta^2\rho_{ii}) }
	{(\bar\theta^2-\theta^2)(\rho_{ii}+\rho_{jj})}
\right)\;.
%=\frac{c_\theta}{\gamma_*}
%  \ln \Big(
%  \frac{e^{\gamma_*\tau/c_\theta}\zeta +\eta}{\zeta+\eta}
%  \Big),
\end{equation}
%where 
%$\zeta := \bar{\theta}^2 \rho_{ii} -\theta^2 \rho_{jj}$  and
%$\eta  := \bar{\theta}^2 \rho_{jj} -\theta^2 \rho_{ii}$.
This is meaningful as long as
$0 \le \tau_{ij}(\theta) \le \tau$, which corresponds to
the condition
\begin{equation}\label{eqn:theta-stop}
\theta^2 / \bar{\theta}^2 \le \rho_{ii}/\rho_{jj} \le \bar\theta^2 / \theta^2\;.
\end{equation}

If $\theta \neq 1/2$ the noise qubit has a unique fixed point,
\begin{equation}\label{eqn:fixp-theta}
\rho_\infty(\theta) = c_\theta\;
	\begin{pmatrix} \bar\theta^2 & 0\\
	  0 & \theta^2 \end{pmatrix}\;.
\end{equation}
Comparing this with the canonical density operator of a qubit with energy level splitting~$\Delta$
at inverse temperature~$\beta~:=~\tfrac{1}{k_B T}$,
\begin{equation}
\rho_\beta := \frac{1}{2 \cosh(\beta \Delta/2)}\begin{pmatrix}
  e^{\beta \Delta/2} &0 \\ 0 & e^{-\beta \Delta/2}\end{pmatrix}\;,
\end{equation}
we can see that the parameter $\theta$ corresponds to the inverse temperature 
\begin{equation}
\label{eqn:AlgCooling-delta}
\beta(\theta) =
\tfrac{2}{\Delta} \operatorname{artanh}\left(\delta(\theta)\right)
\quad \text{with} \quad
\delta(\theta) := \frac{\bar\theta^2-\theta^2}{\bar\theta^2+\theta^2}\;.
\end{equation}
Thus (for $t\to\infty$) the relaxation by the single Lindblad generator $V_\theta$ shares the canonical
fixed point with equilibrating the 
system via the noisy qubit with a local bath of temperature
$\beta(\theta)$.
As limiting cases,  pure amplitude damping is
brought about by a bath with zero temperature $T_\theta=0$ (i.e.~$\theta=0$),
while pure bit-flip shares the canonical fixed point with
the high-temperature limit $T_\theta \to \infty$ (i.e., $\theta \to \tfrac{1}{2}$),
see also Sec.~\ref{sec:heatbath} for the relation to heat baths.

Whereas in a single-qubit system with unitary control and bang-bang switchable noise generator $V_\theta$
it is straightforward to see that
one can (asymptotically) reach all states with purity less or equal to the
larger of the purities of the initial state~$\rho_0$ and~$\rho_\infty(\theta)$,
\begin{equation}
\overline\reach_{\rm 1 qubit,\Sigma_\theta}(\rho_0) = \{\rho\,|\,\rho\prec\rho_0\} \cup \{\rho'\,|\,\rho'\prec\rho_\infty(\theta)\}\;,
\end{equation}
the situation for $n\geq 2$ qubits is more involved:
relaxation of a diagonal state can only be limited to a single pair of
eigenvalues, if all the remaining ones can be arranged in pairs each satisfying
Eqn.~\eqref{eqn:theta-stop}.

However, Eqn.~\eqref{eqn:theta-stop} poses no restriction  in an important special case, i.e.~the task of cooling:
starting from the maximally mixed state, optimal control protocols with period-wise relaxation by $V_\theta$ 
interspersed with unitary permutation of diagonal density operator elements clearly include
the partner-pairing approach~\cite{SMW05} to
\emph{algorithmic cooling} with bias~$\delta(\theta)$ defined in Eqn.~\eqref{eqn:AlgCooling-delta}
as long as $0\leq\theta<{1}/{2}$. Note that this type of algorithmic cooling proceeds also just on
the diagonal elements of the density operator, but it involves no transfers limited to a single pair
of eigenvalues. Let $\rho_\delta$ define the state(s) with highest asymptotic purity achievable by
partner-pairing algorithmic cooling with bias $\delta$.
As the pairing algorithm is just a special case of unitary evolutions plus relaxation brought about by $V_\theta$, 
one arrives at
\begin{equation}
%\overline\reach(\tfrac{1}{2^n}\unity) \supseteq \{\rho\,|\,\rho\prec\rho_\delta\}\;.
%\overline\reach(\rho_0) \supseteq \{\rho\,|\,\rho\prec\rho_\delta\}\quad\text{for any $\rho_0$}\;.
\overline\reach(\rho_0) \supseteq \overline\reach(\rho_\delta) \quad\text{for any $\rho_0$}\;,
\end{equation}
because any state $\rho_0$ can clearly be made diagonal to evolve into a fixed-point state 
obeying Eqn.~\eqref{eqn:theta-stop}, from whence the purest state $\rho_\delta$ can be reached by 
partner-pairing cooling.

To see this in more detail, note that a (diagonal) density operator $\rho_\theta$ of an $n$-qubit system 
is in equilibrium with a bath of inverse temperature $\beta(\theta)$ coupled to its terminal qubit, if the
pairs of consecutive eigenvalues satisfy
\begin{equation}
\frac{\rho_{ii}}{\rho_{i+1,i+1}}=\frac{\bar\theta^2}{\theta^2}\quad\text{for all odd $i < 2^n$}\;.
\end{equation}
Hence (for $\theta\neq1/2$) such a $\rho_\theta$ is indeed a fixed point under uncontrolled drift, i.e.~relaxation by $V_\theta$
and evolution under a diagonal Hamiltonian $H_0$ thus extending Eqn.~\ref{eqn:fixp-theta} to $n$ qubits. 
Now, if (say) the first pair of eigenvalues is inverted by a selective $\pi$~pulse 
(which can readily be realized by unitary controls with relaxation switched off),
a subsequent evolution under the drift term only affects the first pair of eigenvalues as
\begin{equation}
\begin{split}
\frac{1}{\bar\theta^2+\theta^2} \left[\begin{matrix} (\bar\theta^2 + \theta^2  \varepsilon_\theta)
		&\bar\theta^2(1-\varepsilon_\theta)\\
		\theta^2(1-\varepsilon_\theta) &  (\theta^2 + \bar\theta^2  \varepsilon_\theta) \end{matrix}\right]
	\begin{pmatrix} c_\theta \theta^2\\ c_\theta\bar\theta^2\end{pmatrix}
& =\frac{1}{\bar\theta^2+\theta^2} \begin{pmatrix} \bar\theta^2 +   \varepsilon_\theta(\theta^2-\bar\theta^2)\\
		\theta^2 - \varepsilon_\theta(\theta^2-\bar\theta^2) \end{pmatrix}\\[2mm]
& =\left[\begin{matrix}   \varepsilon_\theta & (1-\varepsilon_\theta)\\
		(1-\varepsilon_\theta) &   \varepsilon_\theta \end{matrix}\right] 
	\begin{pmatrix} c_\theta \theta^2\\ c_\theta\bar\theta^2\end{pmatrix} \;.
\end{split}
\end {equation}
%(where both identities involve cancellation of the factor $c_\theta=1/(\bar\theta^2+\theta^2)$). 
In other words, the evolution then acts as a $T$-transform on the first eigenvalue pair. 
Since the switching condition Eqn.~\ref{eqn:theta-stop}
is fulfilled at any time, {\em all} $T$-transformations with  $\varepsilon_\theta\in[0,1]$
on the first pair of eigenvalues can be obtained and preserved during transformations
on subsequent eigenvalue pairs.

Hence from any diagonal fixed-point state $\rho_\theta$ (including $\rho_\delta$ as a special case),
those other diagonal states (and their unitary orbits) can be reached that arise by pairwise $T$-transforms  as long as the
remaining eigenvalues can be arranged such as to fulfill the stopping condition Eqn.~\ref{eqn:theta-stop}.
Suffice this to elucidate why for $n\geq 2$ a fully detailed determination of the asymptotic reachable set in the case
of unitary control plus a single switchable $V_\theta$ on one qubit appears  involved and will thus be
treated elsewhere.

\subsection{Bosonic or Fermionic Heat Baths}
\label{sec:heatbath}

Following the lines of \cite[Ch.~3.4.2]{BreuPetr02},
a qubit with the Hamiltonian
\be
H = - \hbar \omega_0 \frac{\sigma_z}{2}
\ee
coupled to a bosonic (or fermionic) heat bath with inverse
temperature~$\beta$ via a coupling of the form~$\sigma_x \otimes (a +a^\dagger)$
is described in the Born-Markov approximation by a Lindblad equation
(of the form of Eqs.~\eqref{eqn:master}, \eqref{eqn:Lindblad})
with two dissipator terms, $\Gamma_{\sigma^-}$ and~$\Gamma_{\sigma^+}$,
\be
\dot{\rho} = -\big(i \hat H
+\gamma\, (1\pm n(\omega_0))\, \Gamma_{\sigma^-}
+\gamma\; n(\omega_0)\, \Gamma_{\sigma^+}\big)\rho,
\ee
where $n(\omega) := 1 / (e^{\beta \hbar \omega} \mp 1)$ is the Planck
(or Fermi) distribution and $\gamma$~is the relaxation rate constant.
In the $\VEC{}$-superoperator representation
this yields a Liouvillian which takes the form
\be
\Gamma_{T}
\propto
\begin{pmatrix}
0 & 0 & 0 & 1\\
0 & -\frac{1}{2} & 0 & 0\\
0 & 0 & -\frac{1}{2} & 0\\
0 & 0 & 0 & -1
\end{pmatrix}
+n(\omega_0)
\begin{pmatrix}
-1 & 0 & 0 & \pm 1\\
0 & -\frac{1 \pm 1}{2} & 0 & 0\\
0 & 0 & -\frac{1 \pm 1}{2} & 0\\
1 & 0 & 0 & \mp 1
\end{pmatrix}.
\ee
In the zero-temperature limit, $n(\omega_0) \to 0$ and only the $\Gamma_{\sigma^-}$ term remains.
In the limit $T \to \infty$, we have
$n(\omega_0) \to \infty$ for bosons and  
$n(\omega_0) \to \tfrac{1}{2}$ for fermions, and one thus obtains in both cases
\be
\lim_{T \to \infty} \Gamma_T
\propto \Gamma_{\{\sigma^+, \sigma^-\}}
=-
\begin{pmatrix}
-1 & 0 & 0 & 1\\
0 & -1 & 0 & 0\\
0 & 0 & -1 & 0\\
1 & 0 & 0 & -1
\end{pmatrix}.
\ee
This is equivalent to dissipation under the two Lindblad operators
$\{\sigma_x, \sigma_y\}$ since
$\Gamma_{\{\sigma_x, \sigma_y\}} = 2 \Gamma_{\{\sigma^+, \sigma^-\}}$.
In contrast, $\sigma_x$ as the only Lindblad operator (generating bit-flip noise) gives
\be
\Gamma_{\{\sigma_x\}}
=-
\begin{pmatrix}
-1 & 0 & 0 & 1\\
0 & -1 & 1 & 0\\
0 & 1 & -1 & 0\\
1 & 0 & 0 & -1
\end{pmatrix}
\ee
in agreement with Eqn.~\eqref{eqn:GLtheta}.
Note, however, that the propagators $e^{-t \Gamma_\nu}$ generated by $\Gamma_{\{\sigma^+, \sigma^-\}}$ 
{\em vs} $\Gamma_{\{\sigma_x\}}$ 
only act on {\em diagonal} density operators (and those with purely imaginary coherence terms $\rho_{12}=\rho^*_{21}$)  
in an indistinguishable way. However,
the relaxation of the real parts of the coherence terms $\rho_{12}=\rho^*_{21}$ differs: 
$\Gamma_{\sigma_x}$ leaves them invariant, see also Eqn.~\eqref{eqn:expGLtheta},
while $\Gamma_{\{\sigma^+, \sigma^-\}}$ does not.
In other words, $\Gamma_{\sigma_x}$ does have nontrivial
invariant subspaces used in Eqn.~\eqref{eqn:protect},  while $\Gamma_\infty$  does not.
%Hence we don't know if $\Gamma_\infty$ gives a majorization rule
%similar to Theorem 2.
Also for non-zero temperatures, there is no direct equivalence
between relaxation under $\Gamma_T$ and $\Gamma(\theta)$.

%%%%%%%%%%%%%%%%%%
\section{\grape Extended by Noise Controls}
%%%%%%%%%%%%%%%%%%
In state transfer problems the fidelity error function used in~\cite{PRA11} is valid
if the purity remains constant or the target
state is pure. In contrast to closed systems, in open ones these conditions need not hold. 
Thus here we use a full
%scaled
Frobenius-norm distance-based error function instead:
$
\delta_F^2 :=
%\frac{1}{2 |X_\text{target}|_F^2}
\left\|X_{M:0}-X_\text{target}\right\|_F^2,
%= \frac{1}{2} \left(1 +\frac{|X_{M:0}|_F^2}{|X_\text{target}|_F^2}\right)
%-\frac{1}{|X_\text{target}|_F^2}\Re \trace \left(X_\text{target}^\dagger X_{M:0}\right),
$
where $X_{k:0} = X_k \cdots X_1 \text{vec}(\rho_0)$ is the vectorised
state after time slice~$k$,
$X_k = e^{-\Delta t  L_k}$ is the
propagator for time slice~$k$ in the Liouville space, and
$L_k := i\hat H_u(t_k) +\Gamma(t_k)$.
The gradient of the error is obtained as
\begin{align}
\Partial{\delta_F^2}{u_j(t_k)}
&=
%\frac{1}{|X_\text{target}|_F^2}
2 \Re \trace\Big((X_{M:0}-X_\text{target})^\dagger \Partial{X_{M:0}}{u_j(t_k)}\Big),
\end{align}
where
\begin{align}
\Partial{X_{M:0}}{u_j(t_k)}
= X_M \cdots X_{k+1}\Partial{X_k}{u_j(t_k)} X_{k-1}\cdots X_1 \text{vec}(\rho_0).
\end{align}
Exact expressions for the derivatives of~$X_k$~\cite{PRA11}
require $L_k$ to be normal, which does not hold in the general case of
open systems of interest here.
%(i.e.~whenever $[\hat H_u(t_k) ,\Gamma(t_k)] \neq 0$ or a $\Gamma(t_k)$ is
%non-normal).
Instead we may use, e.g., the finite-difference method to compute
the partial derivatives as
\begin{align}
\Partial{X_k}{u_j(t_k)}
&= \lim_{s \to 0} \frac{\exp(-\Delta t (L_k +is \hat{H}_j)) -\exp(-\Delta t L_k)}{s},\\
\Partial{X_k}{\gamma_j(t_k)}
&= \lim_{s \to 0} \frac{\exp(-\Delta t (L_k +s \Gamma_{V_j})) -\exp(-\Delta t L_k)}{s}.
\end{align}
The optimal value of $s$ is obtained as a tradeoff between
the precision of the gradient and numerical accuracy, which starts to
deteriorate when $s$ becomes very small.
%One possible way to choose~$s$ is to pick a random, reasonable point and
%a random direction in the control parameter space, compute the
%approximate derivative at this point along the direction using different
%values of~$s$...

%%%%%%%%%%%%%%
\section{Further Numerical Results}\label{sec:furtherresults}
%%%%%%%%%%%%%%

%%%%%%%%%%%%%%%%%%%%%%%%%%%%%%%%%%%%
\begin{figure}[Ht!]
\hspace{7mm}{\sf (a)}\hspace{40mm}\sf{(b)} \hspace{40mm}{\sf (c)}$\hfill$\\
\includegraphics[width=0.24\columnwidth]{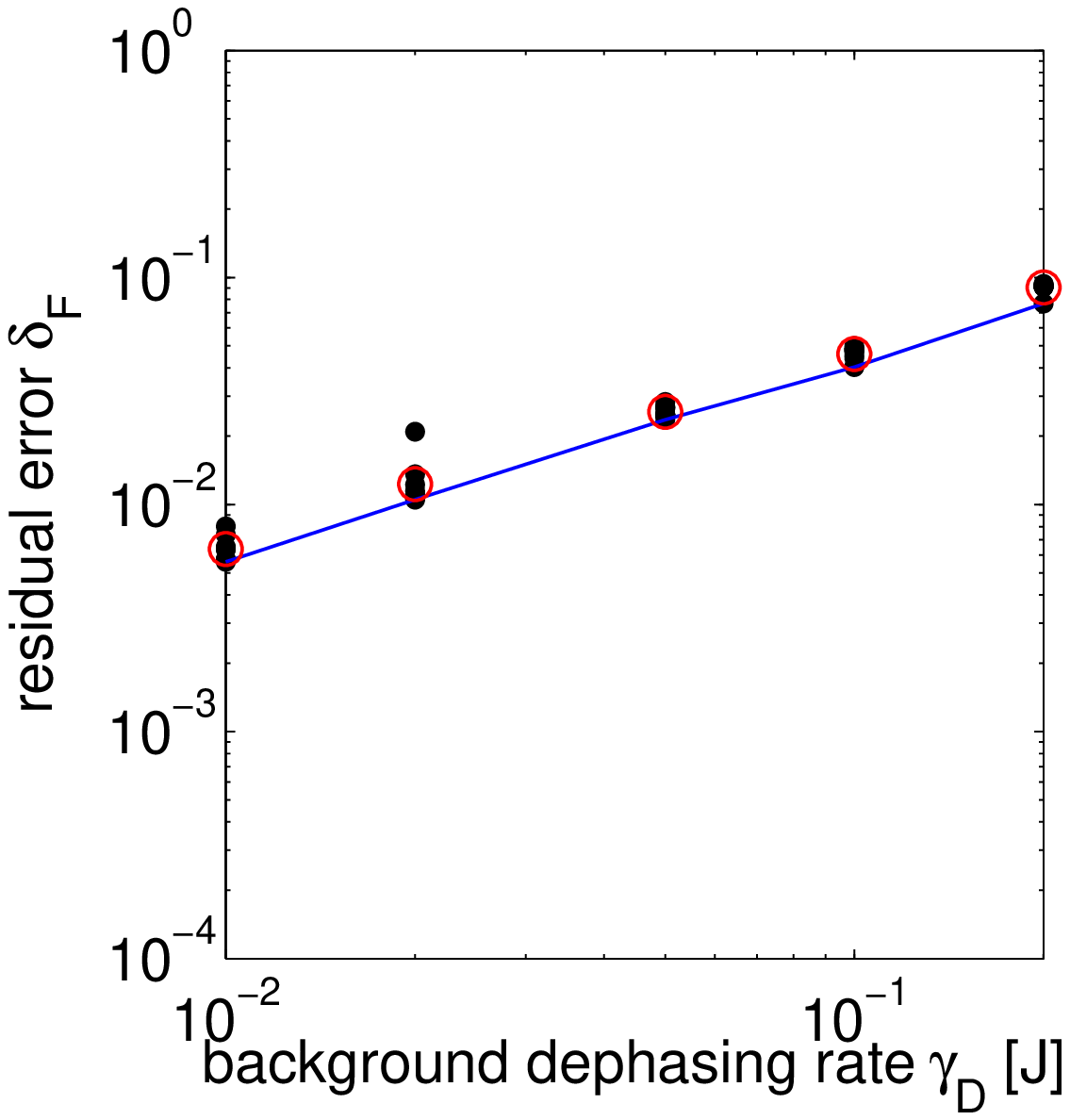}
\includegraphics[width=0.24\columnwidth]{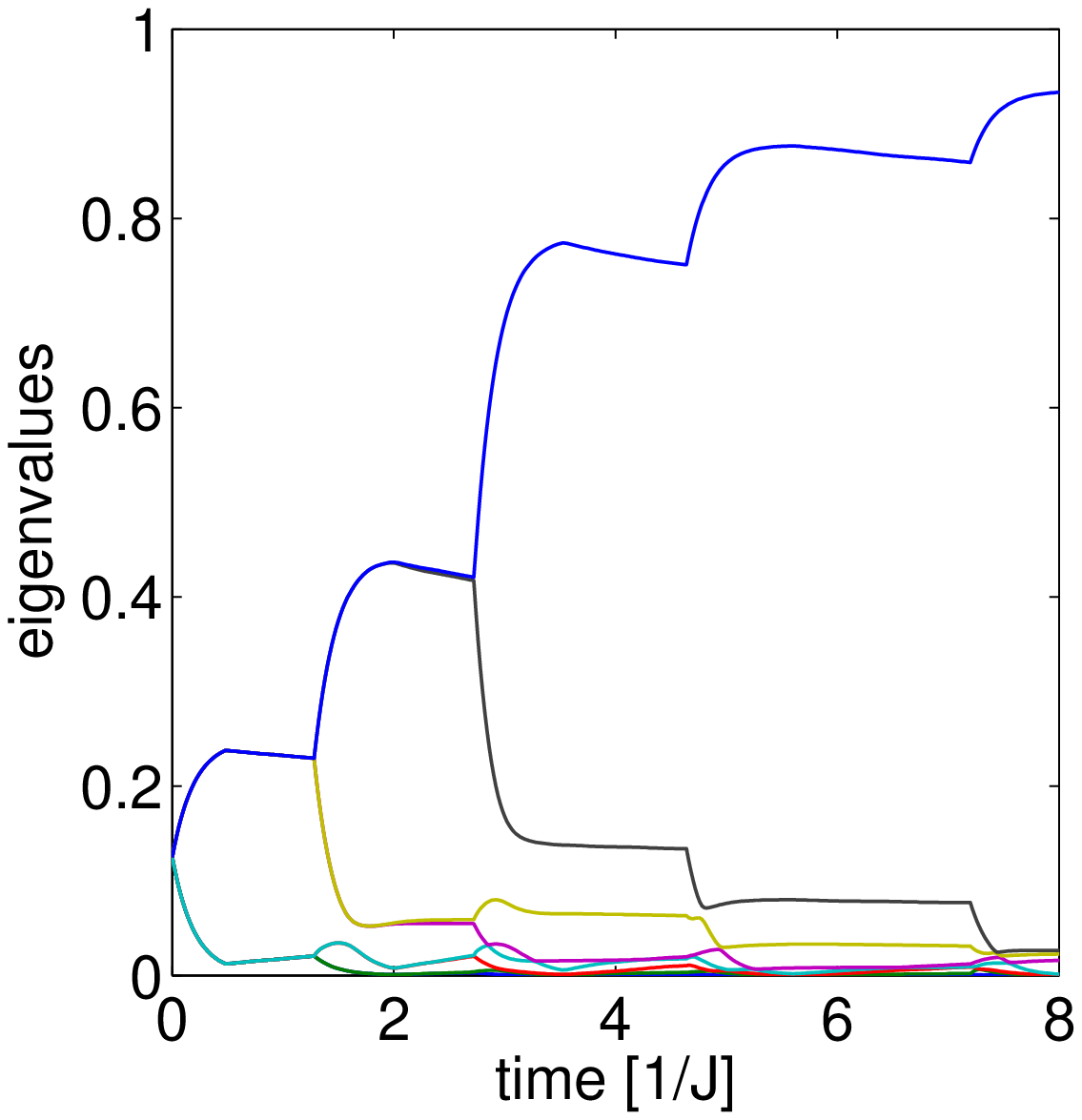}
\includegraphics[width=0.5\columnwidth]{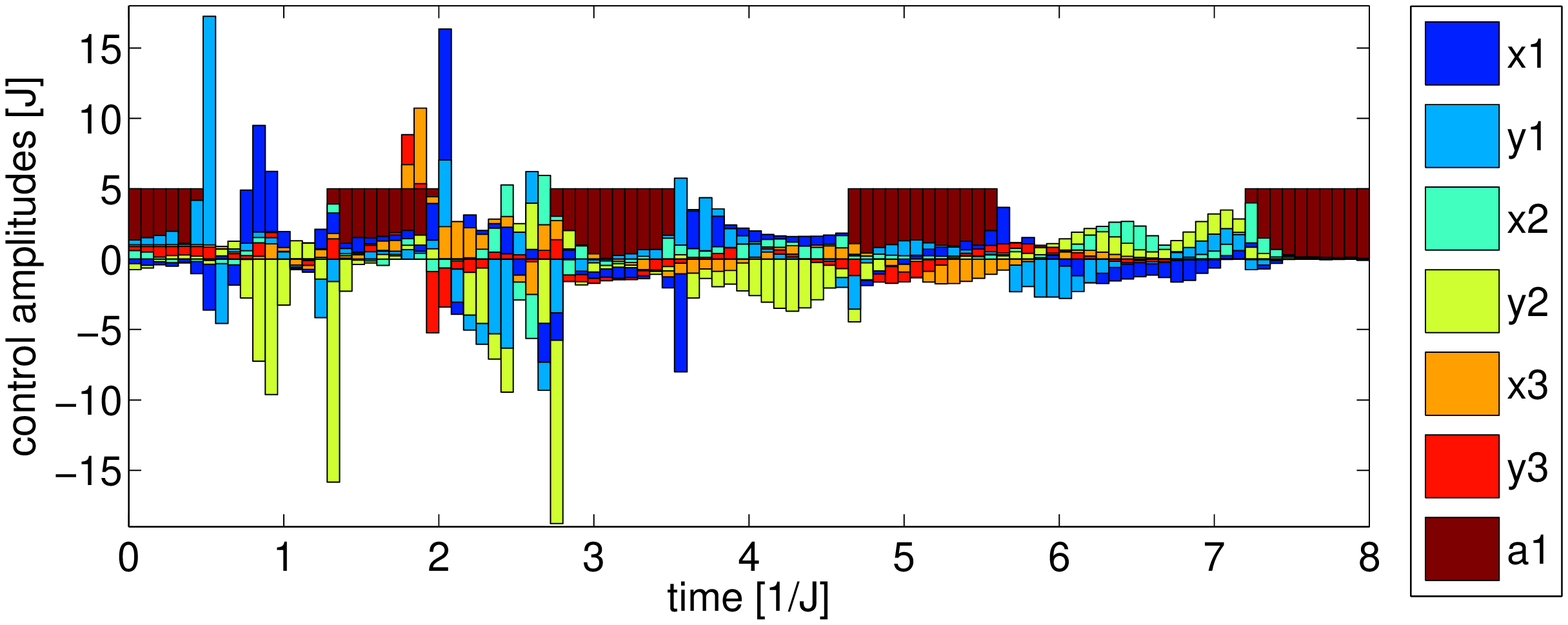}
\caption{\label{fig:th0-dep}
Same as in
Fig.~\ref{fig:th0}, but with additional non-switchable background
dephasing noise on all the three qubits.
(a)~Quality versus dephasing rate~$\gamma_D$ (with $\gamma_D/J\in\{0.01, 0.02, 0.05, 0.1, 0.2\}$)
for sequences of duration~$T=8/J$.
The dots (red circles for averages) are individual numerical optimal-control runs with 
random initial sequences.
(b)~Evolution of the eigenvalues under the best sequence for the strongest background noise ($\gamma_D = 0.2 J$)
leading to the zero-state with a considerably low residual error of $\delta_F\simeq 0.077$.
This sequence (c) shows five relaxative periods with maximal noise amplitude 
on qubit one ($\gamma_{a1}$) for transforming eigenvalues, while 
the unitary actions again mainly take place in the intervals between them.
}
\end{figure}
%%%%%%%%%%%%%%%%%%%%%%%%%%%%%%%%%%%%

{\bf Example~1b.}
Interestingly, the initialisation task of Example~1 can still be accomplished to a good
approximation when unavoidable constant dephasing noise on all the
three qubits is added.
This is shown in Fig.~\ref{fig:th0-dep} for a range of dephasing rate constants reaching from
$1\%$ to $20\%$ of the coupling constant. Though the dephasing does not affect the 
evolution of diagonal states, it interferes with the $i$-swaps needed to
permute the eigenvalues. For $\gamma_D=0.2~J$, numerical optimal control suggests the
sequence Fig.~\ref{fig:th0-dep} (c) with five dissipative steps and increasing time intervals for the
\mbox{$i$-swaps}.

\begin{figure}[Ht!]
\hspace{10mm}{\sf (a)}\hspace{40mm}\sf{(b)} \hspace{40mm}{\sf (c)}$\hfill$\\
\includegraphics[width=0.24\columnwidth]{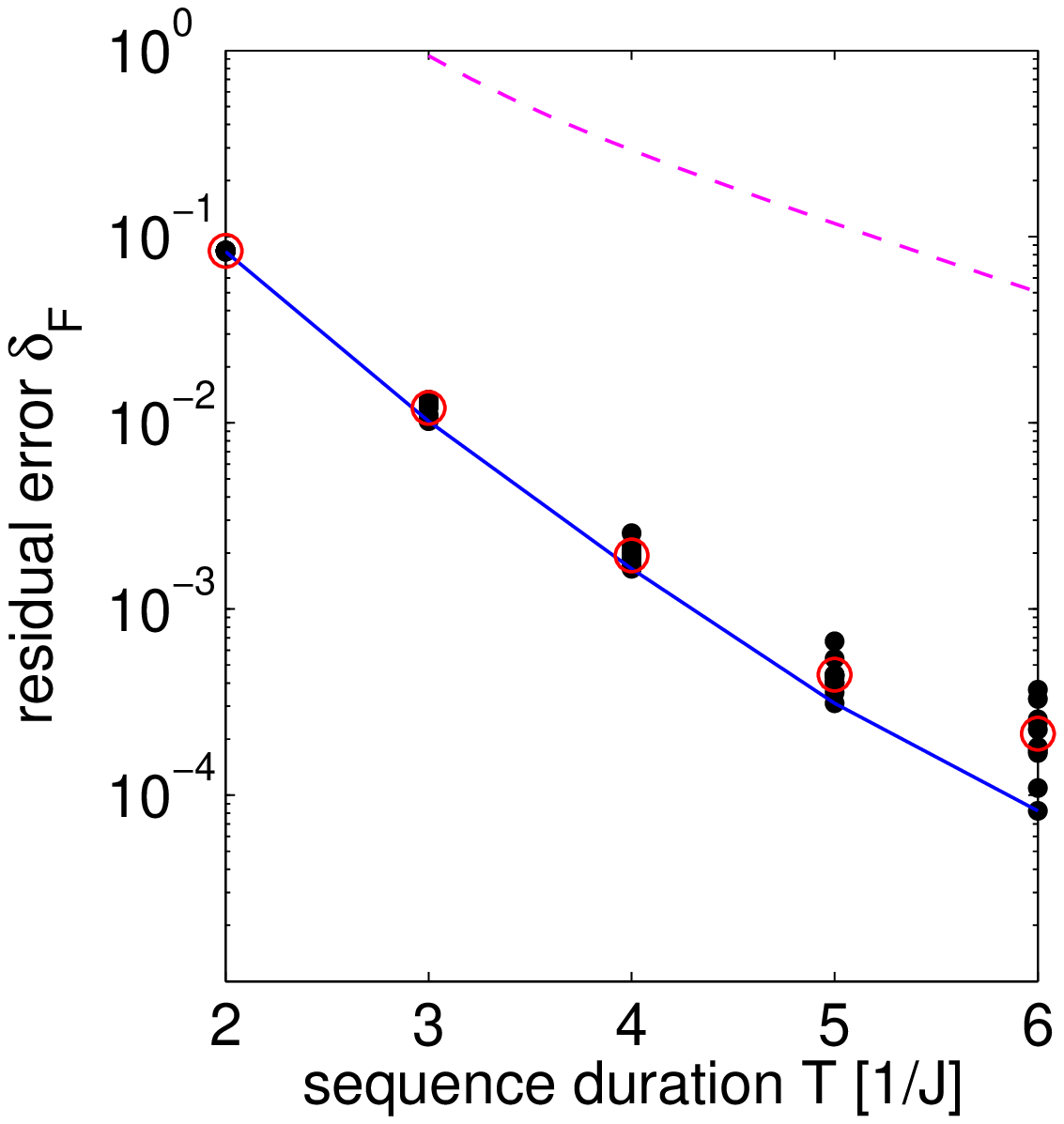}
\includegraphics[width=0.24\columnwidth]{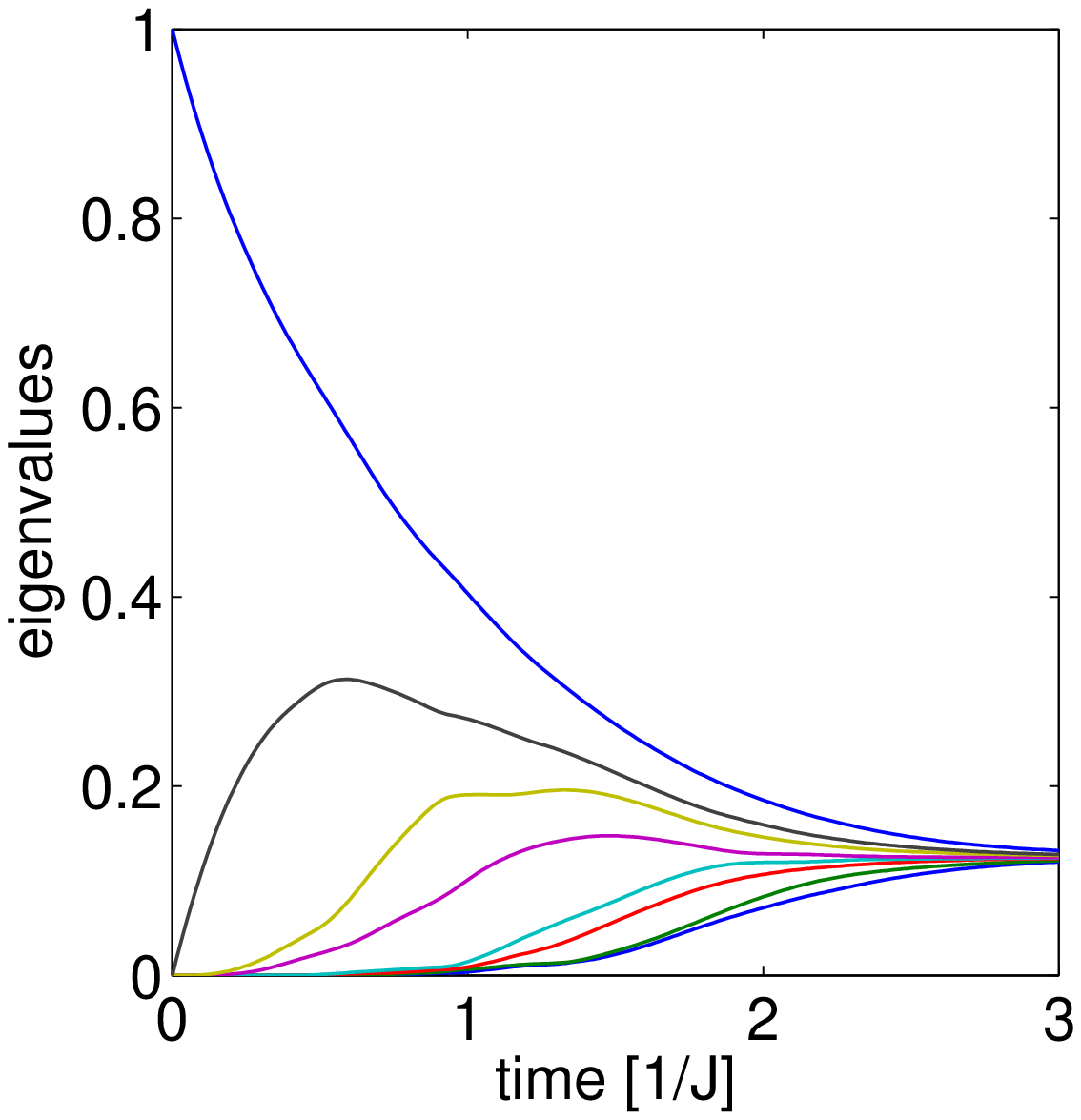}
\includegraphics[width=0.5\columnwidth]{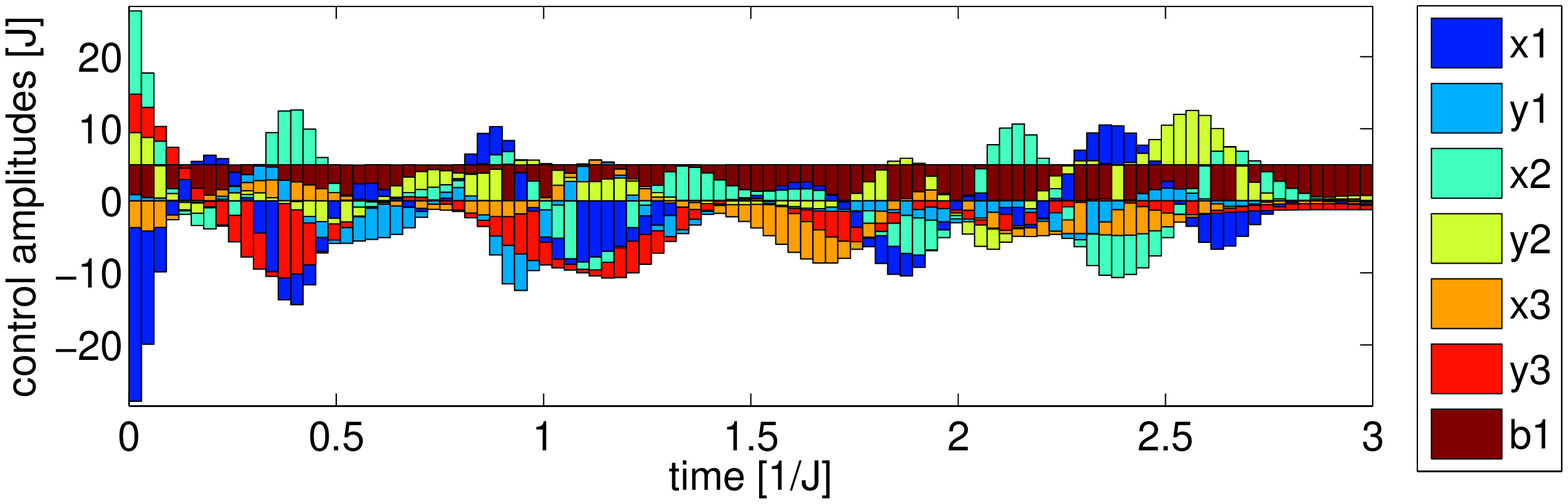}
\caption{\label{fig:0th}
Transfer from the zero-state~$\rho_{\ket{000}}$ to the thermal state~$\rho_\text{th}=\frac{1}{8}\unity$
in a \mbox{3-qubit} Ising-$ZZ$ chain
with controlled bit-flip on qubit one
and local $x,y$-pulse controls on all 
qubits as in {\bf Example~2}.
(a)~Quality versus total duration~$T$, with the dashed line as the upper bound from Eqn.~\eqref{eq:0thest}.
Dots (red circles for averages) denote individual numerical optimal-control runs with 
random initial sequences.
%The dash-dotted line represents a tighter, hypothetical upper bound obtained
%with continuous averaging of the eigenvalues and no separate
%unitaries.
(b)~Evolution of the eigenvalues under the controls of the best of the
$T=3/J$ solutions.
The corresponding control sequence (c) shows that the noise is always
maximised, and the unitary actions
generated by ($u_{x\nu}, u_{y\nu}$) are fully parallelised with it.
}
\end{figure}

\medskip
In {\bf Example~2}, we considered erasing the pure initial
state~$\rho_{\ket{00\ldots0}}$
to the thermal state~$\rho_{\text{th}}$ by controlled bit-flip noise of Eqn.~\eqref{eqn:amp-damp+bit-flip}
to illustrate the scenario of Thm.~\ref{thm:majorisation}.
For $n$~qubits, one may use a similar $n$-step protocol as in Example~1,
this time approximately erasing each qubit to a state proportional to~$\unity$.
%For $n$-qubit Ising-$ZZ$ chains, one may proceed in $n$ steps:  
%on each qubit $q$ let the noise act for the time~$\tau_q$ to 
%approximate a state proportional to~$\unity$ on qubit $q$ and permute the qubits for the next erasure
%on the following qubit.
Again one finds that the residual error $\delta_F$  is minimal for equal~$\tau_q$ to give
\be
\delta_{F_b}^2(\expfactorbitflip)
= \tfrac{1}{2^n} \big( \left(1 +\expfactorbitflip^2\right)^n -1\big),
\ee
where $\expfactorbitflip:= e^{-\gamma_* T_n/(2n)}$.
This yields
\be
\label{eq:0thest}
\begin{split}
T_b &= \binom{n}{2}\tfrac{1}{J} -\tfrac{n}{\gamma_*} \ln \big((2^n \delta_{F_b}^2 +1)^{1/n} -1\big).
\end{split}
\ee
Once again Fig.~\ref{fig:0th}(a) shows that numerical optimal control finds much faster solutions than 
this simplistic protocol.
The noise amplitude tends to be maximised throughout the sequence
with the unitaries fully parallelised, as shown in the example
sequence (c), and reflected in the eigenvalue flow (b).
This works so well because
$\rho_{\text{th}}$ is the unique state majorised by every other state,
and thus all admissible eigenvalue transfers lead towards the goal.

The advantage of optimal-control based erasure becomes evident when 
comparing it to free evolution:
Pure bit-flip noise on one qubit (without coherent controls) would just 
average pairs of eigenvalues once if the free-evolution
Hamiltonian is mere Ising-$ZZ$ coupling, which commutes with the initial state.
Hence free evolution does not come closer to the thermal state than  $\delta_F\simeq 0.61$
and only by allowing for unitary control, erasure becomes feasible for the Ising chain.
% KEEP THESE COMMENT LINES %
%However, if the spin chain is connected by Heisenberg-$XXX$ couplings, this is no
%longer the case, and
%
%after $T=17.5/J$ (with $\gamma_*=5 J$) the same quality is reached as
%after $2/J$ under the controlled scheme for the Ising-$ZZ$ chain, and it takes $T=27/J$ to pass
%$\delta_F\simeq10^{-4}$, which in the controlled case (Fig.~\ref{fig:0th}) is
%accomplished after some $T=3/J$.
%}
%, and it speeds it up
%by a factor of eight for the Heisenberg chain.
%

%%%%%%%%%%%%%%
\section{Analytical Scheme Following Hardy, Littlewood, and P{\'o}lya}\label{sec:HLP}
%%%%%%%%%%%%%%

%%%%%%%%%%%%%%%%%%%%%%%%%%%%%%%%%%%%
%%%%%%%%%%%%%%%%%%%%%%%%%%%%%%%%%%%%

The work of Hardy, Littlewood, and P{\'o}lya~\cite{HLP34} (HLP) provides a {\em constructive}
scheme ensuring the majorisation condition
$\rho_{\rm target}\prec\rho(t)\prec\rho_0\;\text{for all $0\leq t\leq T$}$
to be fullfilled for all intermediate steps. Let
the initial and the target state be given as diagonal vectors with the eigenvalues of the respective density 
operator in descending order, so
$\rho_0=:\diag(y_1, y_2, \dots, y_N)$ and $\rho_{\rm target}=:\diag(x_1, x_2, \dots, x_N)$.
Following~\cite[p32f]{MarshallOlkin}, fix $j$ to be the largest index such that $x_j<y_j$ and let
$k>j$ be the smallest index with $x_k>y_k$. Define $\delta:=\min\{(y_j-x_j),(x_k-y_k)\}$ and
$\lambda:=1-\delta/(y_j-y_k)$. This suffices to construct
\begin{equation}
y':=\lambda y + (1-\lambda) Q_{jk}\; y %[y_1, \dots, y_{j-1},y_k,y_{j+1}, \dots, y_{k-1}, y_j,y_{k+1}, \dots, y_N]
\end{equation}
satisfying $x\prec y' \prec y$. Here the pair-permutation $Q_{jk}$ interchanges the coordinates 
$y_k$ and $y_j$ in $y$.
So  $y'$ is a $T$-transform of $y$, and Ref.~\cite{MarshallOlkin} shows that by $N-1$ successive steps of
$T$-transforming and sorting, $y$ is converted into $x$.
Now the $T$-transforms $\lambda\unity+(1-\lambda)Q_{jk}$
can actually be  brought about by {\em switching on the bit-flip noise} according to 
Eqn.~\ref{eqn:T-trafo} for a time interval of duration
\begin{equation}\label{eqn:tau-jk}
\tau_{jk}:= - \tfrac{2}{\gamma_*}\,\ln |\,1-2\lambda\,|\;.
\end{equation}
%which for the practical purposes shown here could be cut at a maximum of $15/\gamma_*$.

With these stipulations one obtains an iterative analytical scheme for
transferring any $\rho_0$ by unitary control and switchable bit-flip noise on a terminal qubit
%with eigenvalues $\diag(y_1, y_2, \dots, y_N)$ 
into any $\rho_{\rm target}$ satisfying the reachability condition
%with eigenvalues $\diag(x_1, x_2, \dots, x_N)$
$\rho_{\rm target}\prec\rho_0$.

\bigskip

\begin{boxedminipage}[H!]{.95\columnwidth}
{\small 
{\bf Scheme for Transferring Any $n$-Qubit Initial State $\rho_0$ into Any
Target State $\rho_{\rm target}\prec\rho_0$ by Unitary Control and
Switchable Bit-Flip Noise on Terminal Qubit:}
\begin{enumerate}
\item[(0)] switch off noise to $\gamma=0$, diagonalise target $U_x\rho_{\rm target}U^\dagger_x=:\diag(x)$
		to obtain diagonal vector in descending order $x=(x_1, x_2, \dots, x_N)$; keep $U_x$;
\item[(1)] apply unitary evolution to diagonalise $\rho_0$ and set $\tilde{\rho}_0=:\diag(y)$;
\item[(2)] apply unitary evolution to sort $\diag(y)$ in descending order $y=(y_1, y_2, \dots, y_N)$;
\item[(3)] determine index pair $(j,k)$ by the HLP scheme;
\item[(4)] apply unitary evolution to permute entries $(y_1,y_j)$ and $(y_2,y_k)$ of $y$, so $\diag(y)=\diag(y_j,y_k,\dots)$;
\item[(5)] apply unitary evolution $U_{12}$ of Eqn.~\eqref{eqn:protect} to turn $\rho_y=\diag(y)$ into protected state;
\item[(6)] switch on {\bf bit-flip noise on terminal qubit}  $\gamma(t)=\gamma_*$
		for duration $\tau_{jk}$ of Eqn.~\eqref{eqn:tau-jk}
		(while decoupling as in Eqn.~\eqref{eqn:Trotter-dec}) to obtain $\rho_{y'}$;
\item[(7)] to undo step (5), apply inverse unitary evolution $U_{12}^\dagger$ to re-diagonalise $\rho_{y'}$ 
		and obtain next iteration of diagonal vector $y=y'$ and $\rho_y=\diag(y)$;
\item[(8)] go to (2) and terminate after $N-1$ loops ($N:=2^n$);
\item[(9)] apply inverse unitary evolution $U_x^\dagger$ from step (0) to take final $\rho_y$ to 
  $U_x^\dagger \rho_y U_x \simeq \rho_{\rm target}$.
\end{enumerate}
}
\end{boxedminipage}

\medskip

Note that the general HLP scheme need not always be time-optimal:
E.g., a model calculation shows that just the dissipative
intervals for transferring $\diag(1,2,3,\dots, 8)/36$ into $\unity_8/8$
under a bit-flip relaxation-rate constant $\gamma_*=5 J$ and
achieving the target with $\delta_F=9.95 \cdot 10^{-5}$ sum up to $T_{\rm relax}=12/J$ in the HLP-scheme, 
while a greedy alternative can make it within $T'_{\rm relax}=6.4/J$ and a residual error of $\delta_F=6.04\cdot 10^{-5}$.

%%%%%%%%%%%%%%
\section{Outlook on the Relation to Extended Notions of Controllability in Open Quantum Systems}\label{sec:controllabilities}
%%%%%%%%%%%%%%
The current results also pave the way to an outlook on controllability aspects of
open quantum systems on a more general scale, since they are much more intricate than in the case of closed systems 
\cite{VioLloyd01,Alt03,Alt04,Rabitz07b,DHKS08,Yuan09,Yuan11,ODS11,Pechen11,KDH12}.

Here we have taken profit from the fact that like in closed systems 
(where pure-state controllability is strictly weaker than full unitary controllability\cite{AA03,SchiSoLea02a}), 
in open quantum systems
{\em Markovian state transfer} appears less demanding than the operator lift to the most
general scenario of {\em arbitrary quantum map generation} (including non-Markovian ones) first connected to
{\em closed-loop feedback} control in \cite{VioLloyd01}. Therefore in view of experimental implementation,
the question arises how far one can get with {\em open-loop} control including noise modulation and whether the border to
{\em closed-loop feedback} control is drawn by Markovianity. 

\medskip
Due to their divisibility properties \cite{Wolf08a,Wolf08b} that allow for an exponential construction
(of the connected component)  as {\em Lie semigroup} \cite{DHKS08}, {\em Markovian} quantum maps 
are a well-defined special case of the more general completely positive trace-preserving 
(CPTP) semigroup of Kraus maps, which clearly comprise non-Markovian ones, too. 
While some controllability properties of general Kraus-{\em map generation}
have been studied in \cite{VioLloyd01,Rabitz07b}, 
a full account of controllability notions in open systems should also encompass {\em state-transfer} 
to give the following major scenarios:
\begin{enumerate}
\item Markovian state-transfer controllability ({\MSC}),
\item Markovian map controllability (\MMC),
\item general (Kraus-map mediated) state controllability (\KSC) (including the infinite-time limit of \/`dynamic state controllability\/' (\DSC) \cite{Rabitz07b,Pechen11}),
\item general Kraus-map controllability (\KMC) \cite{VioLloyd01,Rabitz07b}. 
\end{enumerate}
Writing \/`$\subseteq$\/' and \/`$\subsetneqq$\/' in some abuse of language for \/`weaker than\/' and
\/`strictly weaker than\/', one obviously has at least
$
\MSC \subseteq \KSC\; \text{and}\; \MMC \subseteq \KMC, 
$
while $\DSC\subsetneqq\KMC$ was already noted in the context of control directly over the Kraus operators \cite{Rabitz07b}.
In pursuing control over environmental degrees of freedom, 
Pechen \cite{Pechen11, Pechen12} also proposed a scheme, where both coherent plus incoherent light 
(the latter with an extensive series of spectral densities depending on ratios over the difference of eigenvalues 
of the density operators to be transferred) 
were shown to suffice for interconverting arbitrary states with non-degenerate eigenvalues in their density-operator
representations.

Yet the situation outlined above is more subtle, since unital and non-unital cases may differ. 
In this work, we have embarked on unital and non-unital Markovian state controllability, 
\MSC\ \footnote{For simplicity, first we only consider
	the extreme case of non-unital maps (such as amplitude damping) allowing for pure-state fixed points and 
           postpone the generalised cases
	parameterised by $\theta$  in Appendix~\ref{app:B} till the very end.}: 

Somewhat surprisingly, in the {\em non-unital} case (equivalent to amplitude damping),
the utterly mild conditions of unitary controllability plus
bang-bang switchable noise amplitude on one single internal qubit (no ancilla) suffice for acting transitively on 
the set of all density operators (Theorem~\ref{thm:transitivity}). Hence these features fulfill the maximal condition $\KSC$ already. 
In other words, for cases of non-unital noise equivalent to amplitude damping (henceforth indexed by \/`{\sf nu}\/'),
$\KSC_{nu}$ implies $\KSC$.
Moreover, under the reasonable assumption that the mild conditions in Theorem~\ref{thm:transitivity} 
are in fact the {\em weakest}
for controlling Markovian state transfer $\MSC_{nu}$ in our context, Theorem~\ref{thm:transitivity} shows that
$\MSC_{nu}$ implies $\KSC$ via $\KSC_{nu}$. 
So in the (extreme) non-unital cases, there is no difference between Markovian and 
non-Markovian state controllability. ---
On the other hand in order to compare non-unital with unital processes, taking  Theorems~\ref{thm:transitivity} and \ref{thm:majorisation} together
proves $\MSC_{u}\subsetneqq\MSC_{nu}$, since the former is restricted by the majorisation condition of 
Theorem~\ref{thm:majorisation}. 

Similarly, in the {\em unital} case (equivalent to bit-flip), the mild conditions of unitary controllability plus
bang-bang switchable noise amplitude on one single internal qubit suffice for achieving {\em all}  state transfers obeying majorisation
(Theorem~\ref{thm:majorisation}). Hence again they fulfill the maximal condition $\KSC_u$ at the same time. 
This is because  state transfer under {\em every} unital CPTP Kraus map (be it Markovian or non-Markovian)  has to meet 
the majorisation condition; so we get $\KSC_u$. On the other hand, the majorisation condition itself imposes the
restriction $\KSC_u\subsetneqq\KSC$.
Again, under the reasonable assumption that the mild conditions in Theorem~\ref{thm:majorisation} 
are in fact the weakest
for controlling Markovian state transfer $\MSC_{u}$ in our context, Theorem~\ref{thm:majorisation} shows that
$\MSC_{u}$ implies $\KSC_u$. Thus also in the unital case, there is no difference between Markovian and 
non-Markovian state controllability.

The results on these two cases, i.e.\ non-unital and unital (in the light of Appendix~\ref{app:B} seen as the limits $\theta=0$ and
$\theta=\tfrac{1}{2}$, respectively), can therefore be summarized as follows:
\begin{corollary}
In the two scenarios of Theorem~\ref{thm:transitivity} (non-unital) and \ref{thm:majorisation} (unital), 
Markovian state controllability already implies Kraus-map mediated state controllability and one finds
\begin{equation}
\begin{array}{c c c c c}
\MSC_{nu} & \Longrightarrow & \KSC_{nu}  & \Longrightarrow & \KSC \\[2mm]
\bigcup\negthickspace\nparallel & &\bigcup\negthickspace\nparallel & 
	\begin{turn}{-45}\raisebox{1.8mm}{$\bigcup\negthickspace\nparallel$}\end{turn} &\\[2mm]
\MSC_u & \Longrightarrow & \KSC_u & &\\ 
\end{array}
\end{equation}
\end{corollary} 
 However, whether $\MSC_\theta\Longrightarrow\KSC_\theta$ also holds in the generalisation of Appendix~\ref{app:B}, where
$\theta$ can range over the entire interval $\theta\in[0,\tfrac{1}{2}]$ (with $\theta=0$ giving the limiting cases
$\MSC_{nu}, \KSC_{nu}$ and $\theta=\tfrac{1}{2}$ yielding $\MSC_u, \KSC_u$), currently remains an open question.

This has an important consequence for experimental implementation of {\em state transfer} in open quantum systems: 
On a general scale in $n$-qubit systems, unitary control plus measurement-based closed-loop feedback from one 
resettable ancilla (as, e.g.,  in Ref.~\cite{BZB11} following \cite{VioLloyd01})
can be replaced by unitary control plus open-loop bang-bang switchable non-unital noise (equivalent to amplitude damping)
on a single internal qubit. This is because both scenarios are sufficient to ensure Markovian and non-Markovian
state controllability \KSC. {\bf Example~5} in the main part illustrates this general simplifying feature.

\medskip

Yet some questions with regard to the operator lift to {\em map synthesis} remain open:
Assessing a demarcation
between \MMC and \KMC (and their unital versus non-unital variants) seems to require 
different proof techniques than used here. 
In a follow-up study we will therefore further develop our lines of assessing the differential geometry 
of Lie semigroups in terms and their Lie wedges \cite{DHKS08,ODS11} to this end,
since judging upon Markovianity on the level of Kraus maps is known to be more intricate \cite{Wolf08b,Wolf12}.
More precisely,
{\em time-dependent Markovian} channels come with a general form of a Lie wedge
in contrast to {\em time-independent Markovian} channels, whose generators form the special structure of a
Lie semialgebra (i.e.\ a Lie wedge closed under Baker-Campbell-Hausdorff multiplication). In  \cite{DHKS08},
we have therefore drawn a detailed connection between these differential properties of Lie semigroups and the 
different notions of divisibility studied as a defining property of Markovianity in the seminal work \cite{Wolf08a}.

Again, these distinctions will decide on simplest experimental implementations in the sense
that measurement-based {\em closed-loop feedback} control may be required for non-Markovian maps in \KMC, 
while {\em open-loop} noise-extended control may suffice for Markovian maps in \MMC.
More precisely, 
{\em closed-loop feedback} control was already shown to be {\em sufficient} for \KMC  in \cite{VioLloyd01}
(which was the aim that work set out for), yet it remains to be
seen whether it is also {\em necessary}, and in particular, if it is necessary for the (supposedly) weaker notion \MMC.
If it turns out {\em not} to be necessary, then  measurement-based closed-loop feedback control on a system extended by one 
resettable ancilla \cite{VioLloyd01,BZB11,SMB12} would be not be stronger than our open-loop scenario of full unitary control extended by 
(non-unital) noise modulation not only in the case of state transfer, but also in quantum-map synthesis with direct
bearing on the simplification of quantum simulation experiments \cite{SMB12}.

%\bibliographystyleA{apsrev4-1}
%\bibliographyA{control21ville}

%%%
%%%%%%%%%%
\end{document}